\documentclass[pdflatex,sn-mathphys]{sn-jnl}

\jyear{2021}%
\usepackage{dot2texi}
\usepackage{algorithm}
\usepackage{algpseudocode}
\theoremstyle{thmstyleone}%
\theoremstyle{thmstyletwo}%

\theoremstyle{thmstylethree}%
\usepackage{graphicx}
\usepackage{float}
\usepackage{caption}
\usepackage{subcaption}

\newcommand{\etal}{{\em et al.}}
\long\def\omitit#1{}
\begin{document}

\title[Activity-preserving time dilation]{Exploring the effects of activity-preserving time dilation on the dynamic interplay of  airborne contagion processes and temporal networks using an interaction-driven model}


\author[1]{\fnm{Alex} \sur{Abbey}}\email{alexzabbey@gmail.com}
\equalcont{These authors contributed equally to this work.}

\author[1]{\fnm{Yanir} \sur{Marmor}}\email{yanirmr@gmail.com}
\equalcont{These authors contributed equally to this work.}

\author[2]{\fnm{Yuval} \sur{Shahar}}\email{yshahar@bgu.ac.il}

\author*[1]{\fnm{Osnat} \sur{Mokryn}}\email{ossimo@gmail.com}

\affil[1]{\orgdiv{Information Systems}, \orgname{University of Haifa}, \orgaddress{ \city{Haifa},   \country{Israel}}}

\affil[2]{\orgdiv{Software and Information Systems Engineering}, \orgname{Ben Gurion University}, \orgaddress{ \city{Beer Sheva},  \country{Israel}}}


\abstract{Contacts' temporal ordering and dynamics are crucial for understanding the transmission of infectious diseases.   
We introduce an interaction-driven model of an airborne disease over contact networks. We demonstrate our interaction-driven contagion model, instantiated for COVID-19, over history-maintaining random temporal networks and real-world contacts. We use it to evaluate temporal, spatiotemporal, and spatial social distancing policies. We find that a spatial distancing policy is mainly beneficial at the early stages of a disease. 

We then continue to evaluate temporal social distancing, that is, timeline dilation that maintains the activity potential.  
We expand our model to consider the exposure to viral load, which we correlate with meetings' duration. Using real-life contact data, we demonstrate the beneficial effect of timeline dilation on overall infection rates. 
 Our results demonstrate that given the same transmission level, there is a decrease in the disease's infection rate and overall prevalence under timeline dilation conditions.  We further show that slow-spreading pathogens (i.e., require more prolonged exposure to infect) spread roughly at the same rate as fast-spreading ones in highly active communities. This is surprising since slower pathogens follow paths that include longer meetings, while faster pathogens can potentially follow paths that include shorter meetings, which are more common.  Our results demonstrate that the temporal dynamics of a community have a more significant effect on the spread of the disease than the characteristics of the spreading processes.}

\keywords{COVID-19, Spatial pods, Temporal pods, Temporal activity, Infection rate}



\maketitle

\section{Introduction}\label{sec1}
The SARS-CoV-2 coronavirus disease 2019 (COVID-19) has created a global crisis.  In response, governments and local authorities employed Non-Pharmaceutical Interventions (NPIs)~\cite{anderson2020will} to limit mobility, social interactions, and gathering to slow - or even contain - the spread of the virus~\cite{courtemanche2020strong,flaxman2020estimating,ferguson2006strategies}.

Here we present a realistic interaction-driven airborne disease model adapted to COVID-19. The interaction-driven temporal contagion model allows susceptible-exposed-infectious-removed (SEIR) node states and computes the probability of infection at the end of each time window, thereby allowing a latent exposed state during that time window.

We demonstrate its performance over temporal random networks and real-world interactions and use it to investigate the effect of temporal, spatial, and spatiotemporal distancing policies on the infection rate. We examine the effect of the NPIs both in the early days of the disease and when it is more prevalent in the community and evaluate them for different pathogens of the disease. 
 
One social distancing solution is the creation of spatial distancing pods. In Israel, for example, to avoid school lockdowns~\cite{donohue2020covid}, social distancing pods were employed in schools~\cite{heymann2004influence}. Each class was split into two disjoint sub-groups, each considered a separate pod. Due to a lack of staff and infrastructure,  each class's pods arrived at school on alternate days, thus creating both a temporal and spatial division. 

Intuitively, separating people into smaller groups, as is the case with spatial pods, reduces the probability of infection as the number of different people exposed decreases. However, the case with temporal pods is different. In temporal pods, one maintains all meetings, albeit on a longer time scale. It is closely related to individual activity potential, which was shown to be virtually independent of the time scale over which the activity potential is measured~\cite{perra2012activity}. Hence, it may be expected that reducing the per-day meetings while maintaining the overall activity would not change the infection rate as scheduling all needed meetings might prolong the number of days people interact (for example, lengthening a conference to allow all planned meetings to occur). On its face value, one might conjecture that increasing the number of days people meet might not necessarily decrease and might even increase the probability of getting infected. Thus, it is not clear that temporal pods would be a good strategy to decrease the spread of the disease.  

To evaluate the Israeli schools distancing policy, we compared the combined scenario of temporal and spatial pods with a random one, where only a randomly selected subset of students attend school each day. We evaluate a spatiotemporal scenario to a temporal one by executing our interaction-driven model over temporal random networks created to simulate the two strategies.

Our interaction-driven model preserves the order and timing of contact interactions, previously demonstrated as a key to accurate modeling~\cite{holme2012temporal,delvenne2015diffusion,masuda2017introduction}. To generate temporal random networks with temporal ordering, we implemented the algorithm suggested by Zhang \etal~in~\cite{zhang2017random} that generates a series of temporal random networks with continuous-time network histories. We set the temporal density in these networks to be stationary to avoid the average fallacy. 

Our results demonstrate that prolonging the timeline by alternating days while splitting the students into two disjoint groups lowers the infection rate. However, these results are more definite at the early stages of the disease, with a low number of initial patients. When examined with a larger number of initial patients, the disjoint groups' added benefit is mitigated. We further validate this result in a follow-up experiment using only spatial pods, demonstrating that spatial, social distancing NPIs are more beneficial in the disease's early days.  

The spreading of contagious respiratory diseases, such as severe acute respiratory syndrome (SARS) and COVID-19, is an exposure-based process that correlates with the duration of the exposure~\cite{rea2007duration,smieszek2009mechanistic,stehle2011simulation,nagel2021realistic,echeverria2021estimating}. 
To incorporate this aspect, we expand our interaction-based model to include the \textit{duration of interactions} and consider this duration as \textit{correlated with the transmitted viral load}. This enables us to create a probabilistic model that does not equally consider interactions of different lengths and allows us to follow the spread of the disease on a real contact network while considering the duration of encounters. 

To model the disease progression in a real-life community, we use the real-world encounters data from the Copenhagen Networks Study (CNS) dataset~\cite{stopczynski2014measuring,Stopczynski:2015aa,sapiezynski2019interaction}. The advantages of working with real-life temporal interactions are numerous, as human interactions are bursty, temporal, highly contextual, and networked~\cite{barabasi2005origin,holme2012temporal,mokryn2016role,fumanelli2012inferring,pastor2015epidemic,delvenne2015diffusion}. The CNS dataset contains, on top of the contact information, the duration of the encounters. 

We then continue examining the effect of temporal pods, i.e., timeline dilation while preserving the activity potential, on contagion in real-world networks while considering meetings' duration. We split each day into two or three time windows. Each time window is diluted meetings-wise. Thus, while the overall activity potential remains the same, the per time window amount of meetings changes. Our results show that using temporal pods, when the spatial order and activity potential are maintained by lengthening the number of days people meet, reduces the spread of the disease.  

The results are more surprising when evaluated for different pathogens by considering pathogens with different minimal exposure duration needed for contagion~\cite{rea2007duration,li2007role,noakes2006modelling,sze2010review,ferretti2020quantifying,luo2021infection,teyssou2021delta}. Overall, we find that temporal pods lower the ability of pathogens to infect, as infectious people have fewer opportunities to infect before they are removed from the network. 
We further show that the temporal dynamics of the network dominate the spreading of the disease and that different pathogens spread similarly in networks with a relatively large number of highly active agents.

Our results indicate that the temporal dynamics of the network have a more significant effect on the spreading of the disease than changes in the spreading process characteristics that reduce the exposure time. We further show that lowering the temporal dynamics can significantly reduce the disease's spread. This finding is important for modeling networks and has implications for analyzing epidemic spreading and viral processes propagation in networks.

\section{Methods}
\subsection*{Temporal random networks modeling}
Zhang et al. ~\cite{zhang2017random} suggest a temporal random network with changing dynamics that follow a Markov process,  allowing for a continuous-time network history (see Supplementary Section 1.1 for details). 
Using this model allows for a fundamental unit of analysis, the entire network history while keeping a stationary density. Further, we chose the parameters s.t. to allow for stationary density (see Supplementary Section 1.1 for details). An impediment to modeling contagious disease over temporal random networks is the fallacy of the average, which allow for various sequences of graphs to have similar average density (See Supplementary Section 1.1). Hence, we require that each graph in the network is created with the same temporal density, thus creating sequences of graphs with stationary density.

To ensure that no anomalies in one graph created an unusual effect, we used a mix of randomly generated graphs. We then conducted a Two-sample Kolmogorov–Smirnov test on each pair of graphs and determined that for each couple, we cannot reject the null hypothesis that both were drawn from the same probability distribution, with 95\% significance~\cite{miller2020size}. 

\subsection*{A temporal interaction-driven contagion model }

At each time window $T_\tau$ with $K_n$ interacting nodes, the probability of a node $i, i \in K_n$ to become exposed is calculated as the complement of the probability of not being exposed in any of the encounters during that time window with infectious nodes, as follows: 
\begin{equation}
    P_i(S \rightarrow E)= 1-\prod^{N_i^\tau}(1- P_{\text{max}})
    \label{eq:exp} 
\end{equation}
Where $N_i^\tau$ is the subset of infected nodes in time window $T_\tau$ that interacted with node $i$ during that time and thus might potentially expose it to the infection, and $P_{\text{max}}$ is the probability of being infected during a maximal exposure. (For example, even on relatively isolated, dense sites such as the Princess Diamond ship during the first wave of the COVID-19 pandemia, and with an air conditioning system that might well have  distributed the virus to multiple cabins, not more than 20\% of the passengers and personnel were infected. 
In other such cases, the maximal infection probability seemed to lie anywhere between 20\%  and 60\%).  

Further details of the per-node state machine of the SEIR-like model appear in Supplementary Section 1.3.

\subsection*{Real-life encounters: considering encounters duration heterogeneity}
\label{sec:rle}

We  model the CNS social network of interactions (see Supplementary Section 1.2) $\Gamma$ as a sequence of $L$ consecutive undirected weighted temporal graphs $\{G_\tau \in \Gamma, \tau \in L\}$ where each temporal snapshot graph $G_\tau=(V_\tau,E_\tau)$ denotes the subset of interacting nodes $V_\tau$ during the $\tau$ temporal window and the weighted edges $E_\tau$ the interactions during this time. 
Each edge is a  distinct interaction. Edge weight corresponds to the {\em duration} of the interaction (measured in seconds or minutes). 

\subsubsection*{Temporal interaction-driven contagion model with various exposure levels}
\label{sec:tm-el}
We model the probability of infection at each interaction with an infected individual as a Sigmoid function of the length of the interaction, where after crossing a lower threshold of minimal time for infection, the probability increases exponentially with the duration up to a maximal duration after which the probability stays stable. This is inline with epidemic understanding of a SARS-like airborne disease~\cite{rea2007duration,ferretti2020quantifying}.

In practice, at each encounter during a time window $\tau$, there is a probability for a node $i$ to get exposed and infected that is calculated as follows.  Let $d_{i,k}$ be a non-zero value for the strength of edges that enter the focal node $i$ from {\em infected node k}, where $k \in K$, the set of infected nodes that $i$ encounters.
 
\begin{equation}
    \forall k \in K,  d_{i,k}=
    \begin{cases}
    d_{i,k} & d_{i,k} \geq D_{\text{min}}\\
    P_{\epsilon} &  d_{i,k} < D_{\text{min}}  
    \end{cases}
\end{equation}
    Where $D_{\text{min}}$  is the minimum exposure required for infection, i.e., the minimal interaction duration needed for infection. If the interaction is shorter than $D_{\text{min}}$, we set the edge's strength to a minimal probability that will not zero the equation but reduce the probability of being infected due to this encounter.
    
    Each time-window, $\tau$,  the probability of node $i$ to get infected as the complement of the probability of not being infected in any of its encounters with infectious nodes in that time window:
\begin{equation}
    P_i(S \rightarrow E,\tau)= 1-\prod_{k}^K(1-\min(\frac{d_{i,k}}{D_{\text{max}}}, 1)\cdot P_{\text{max}})
    \label{eq:vl2} 
\end{equation}
Where $P_i(S \rightarrow E,\tau)$ is the probability of node $i$ in state {\em Susceptible} to transition from state {\em Susceptible} to state {\em Exposed}; $d_{i,k}$ is the duration of the $k_{th}$ interaction of node i; $P_{\text{max}}$ is the probability of being infected given a maximal [i.e., continuous] exposure to the infecting agent; and $D_{\text{max}}$ is a normalization factor that denotes the [minimal] duration of the exposure for which the probability of infection is maximal.   $P_{\text{max}}$ represents, for example, the fact that some of the population might  already be protected for various reasons against the infecting agent, and thus, typically, $P_{\text{max}} < 1$.

The literature further shows that COVID-19 variants are associated with different exposure levels of transmitted viral load~\cite{luo2021infection,teyssou2021delta}. Thus, the virality of such pathogens is correlated in our modeling with the minimum exposure latency, $D_{\text{min}}$. We use it to determine the effect of density on virality by measuring the virality of variants that correspond to various $D_{\text{min}}$ values.

\section*{Results}

\subsection*{Using the interaction-driven model to evaluate NPIs}
In this experiment, we model what we referred to as the Israeli schools' distancing strategy. According to that policy, the classes were split into two disjoint groups, creating spatial social distancing pods. Then, the groups attended schools on alternate days, creating in addition temporal pods. 

In our experiment, we first generated random networks with $10,000$ time windows of $1000$ nodes each. Each time window consists of $288 \times 5\text{-minute}$ intervals, as in a day. Meetings occur during these intervals. The continuous-time network history requirement of the random algorithm defines a probability for the appearance of new edges and the removal of edges. It is computed per each day time window. 
To simulate the schools' policy, we split the nodes into two disjoint groups of $500$ nodes each. Each such group was active on alternate time windows. To evaluate the disease progression, we simulated our interaction-driven contagion algorithm over the resulting network with two initial patients randomly chosen in one of the first two time windows (days). This scenario was then compared to a random one, in which at each time window (day), a randomly chosen group of $500$ nodes is active.

\begin{figure}[!ht] 
  \begin{subfigure}[b]{\linewidth}
    \centering
    \includegraphics[width=0.97\linewidth]{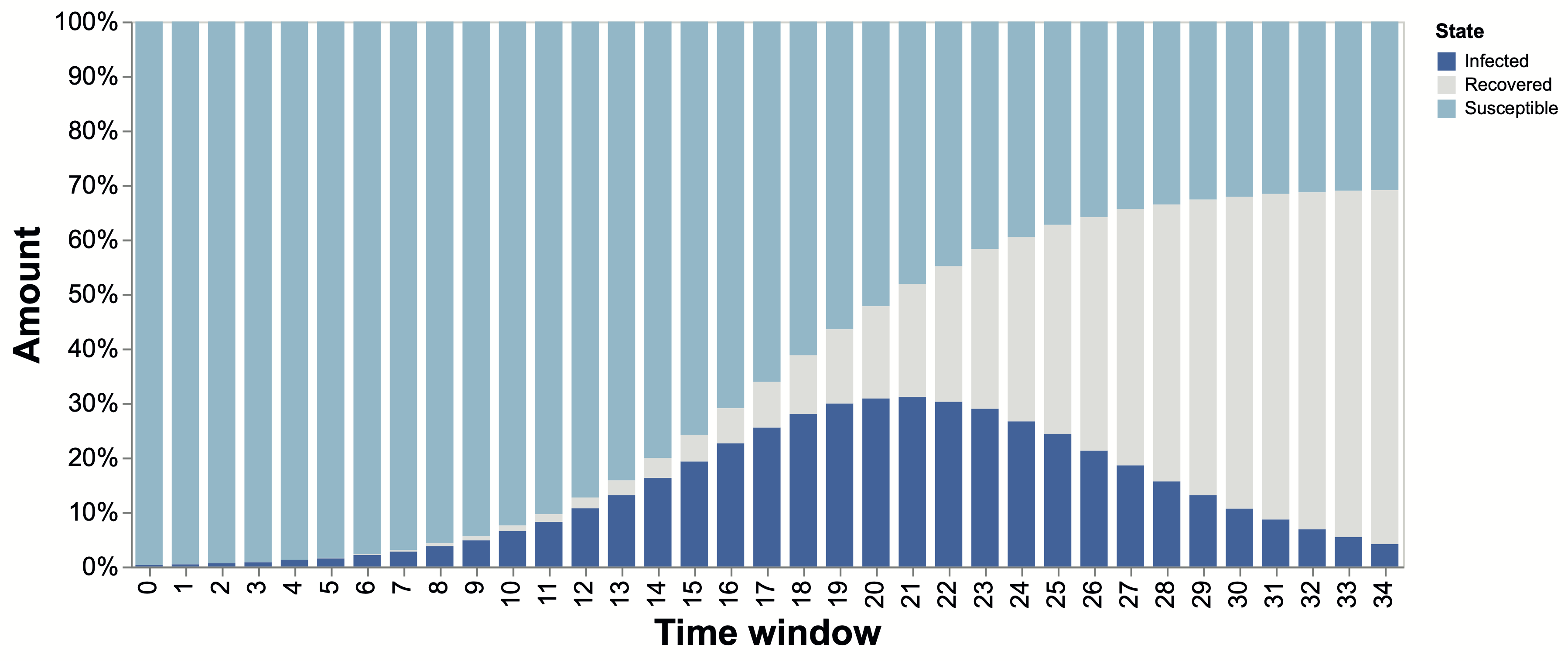} 
    \caption{Temporal and spatial pods, total infection $ = 69.060\% $} 
      \label{fig:f11}
  \end{subfigure} 
    \begin{subfigure}[b]{\linewidth}
    \centering
    \includegraphics[width=0.97\linewidth]{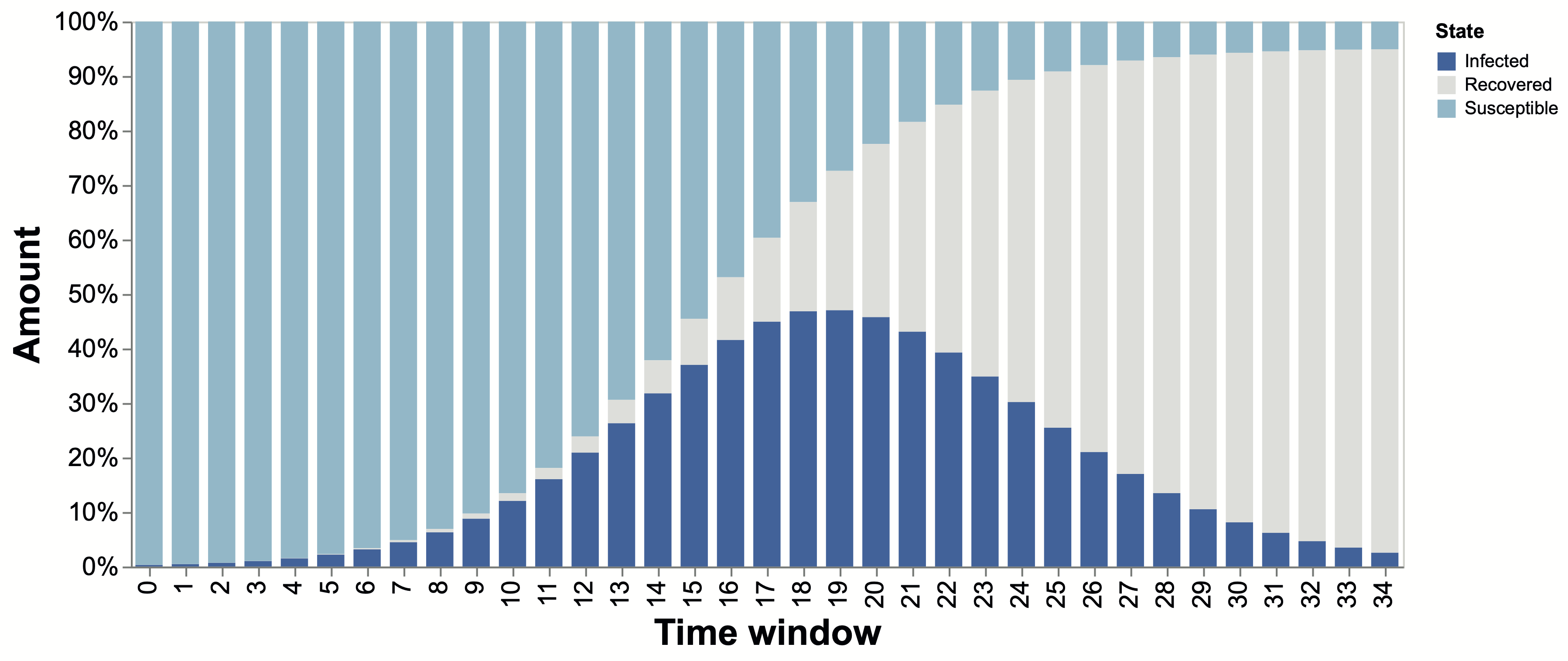} 
    \caption{Random temporal pods, total infection = $94.91\%$} 
    \label{fig:f12} 
  \end{subfigure}

  \caption{Evaluation of the Israeli schools distancing policy. Using our interaction-driven model, we follow the progress of the disease over temporal random graphs with $1000$ nodes each. Each graph consists of $288 \times 5\text{-minute}$ intervals. Panel (a) shows the progress of the disease for a scenario in which two disjoint groups of $500$ nodes each that are active on alternating days. Panel (b) shows the progress of the disease for a scenario in which a randomly selected subset of $500$ nodes is active every day. Each point is the result of 1000 iterations. Error bars are omitted as they are negligible (See Supplementary Section S21 for comparing error bars across a small number of iterations and the current 1000 iterations experiment). Two initial patients are chosen randomly from one of the first two days. The overall infection rate is higher in the case of a random selection of nodes, depicted on panel (b).}
  \label{fig1}
\end{figure}

Figure~\ref{fig1} depicts the results of the experiment. In the upper panel, Figure~\ref{fig:f11}, we see the progression of the disease for the simulated Israeli schools' policy, while Figure~\ref{fig:f12} depicts the progression of the disease for the randomly chosen subset of $500$ nodes each day. The policy that required both the separation of groups and the arrival on alternate days indeed lowers the total infection significantly, as about 70\% were infected compared with almost 95\% in the random scenario. 

As used in the above experiment, two initial patients in a network of $1000$ nodes would correspond to the situation in a community in the early days of the disease. We continue to test the effectiveness of the policy for schools in times when the disease is more widespread.

\begin{figure}[!ht] 
      \begin{subfigure}[b]{\linewidth}
    \centering\includegraphics[width=0.97\linewidth]{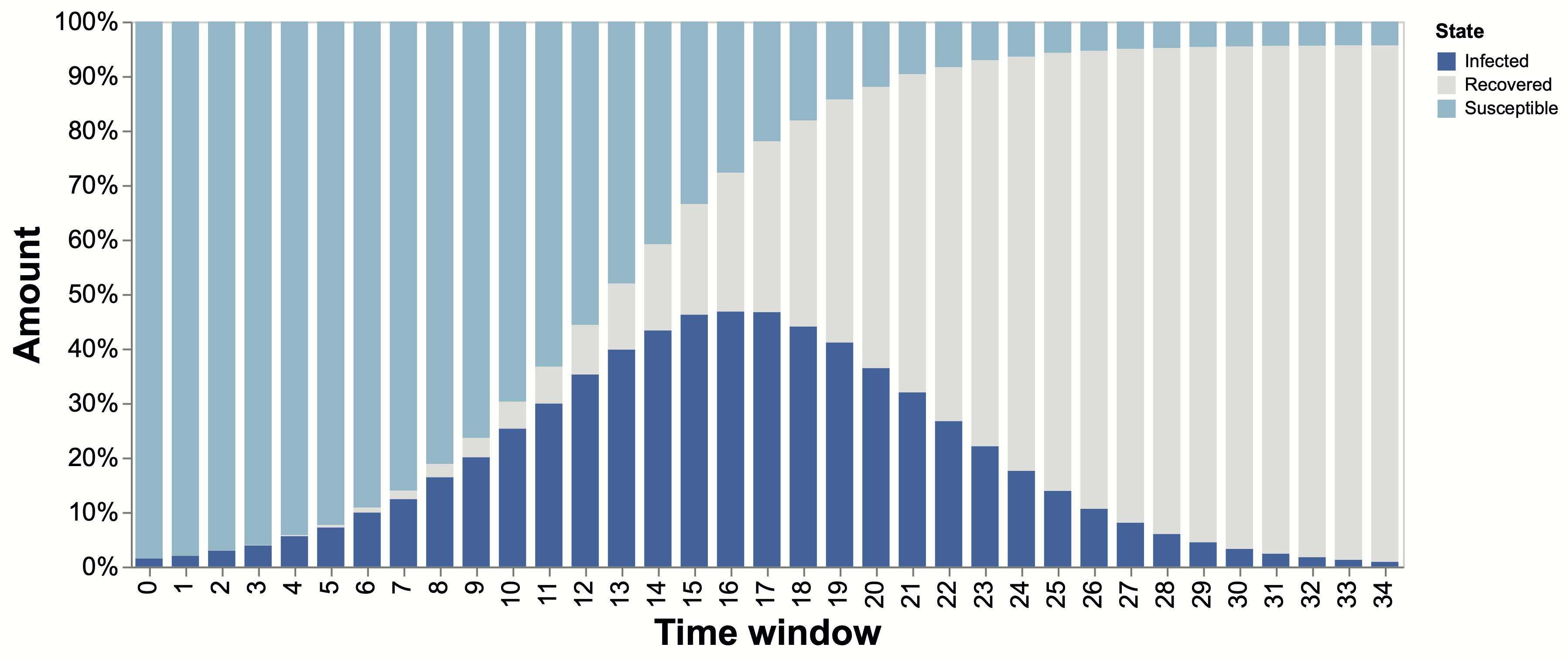} 
    \caption{Temporal and spatial pods, total infection = $95.62\%$ } 
    \label{fig:f13} 
  \end{subfigure}
 \begin{subfigure}[b]{\linewidth}
   \centering
    \includegraphics[width=0.97\linewidth]{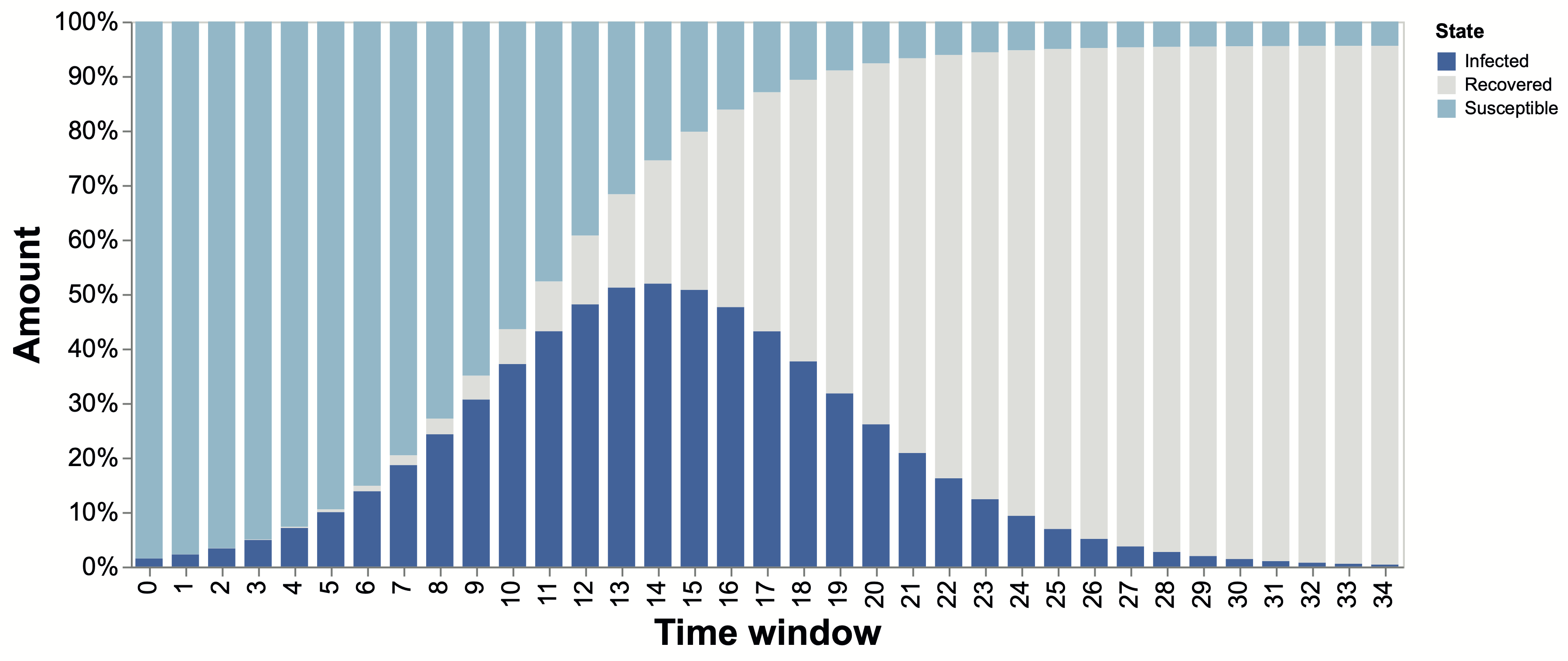} 
    \caption{Temporal and spatial pods, total infection = $95.54\%$ } 
    \label{fig:f14} 
  \end{subfigure} 
  \caption{The effectiveness of the policy for schools when the disease is more prevalent in a community. The possibility of encountering an infectious person is higher,  hence we increase the number of initial patients to ten.  The experiment was performed over our generated random networks with $10,000$ time windows of $1000$ nodes each. Temporal pods contain 500 nodes each, chosen either as disjoint groups (a) or randomly (b). Each point in the graph is the result of 1000 iterations. Error bars are omitted (See Supplementary Section S21). The effect of adding a layer of spatial pods on top of the temporal pods is mitigated when the disease is more prevalent in the population.}
  \label{fig:tenp} 
\end{figure}

Figure~\ref{fig:tenp} shows the disease progression in the same scenarios, albeit with ten initial patients. Each initial patient is chosen randomly from one of the first two days. We see that the total infection rate is roughly the same in both cases examined. This may imply that social distancing pods (spatial pods) of this sort can be effective during the first days of a disease but are less effective when the disease is more prevalent in the community. 

\begin{figure}[!ht] 
  \begin{subfigure}[b]{0.48\linewidth}
    \centering
    \includegraphics[width=0.75\linewidth]{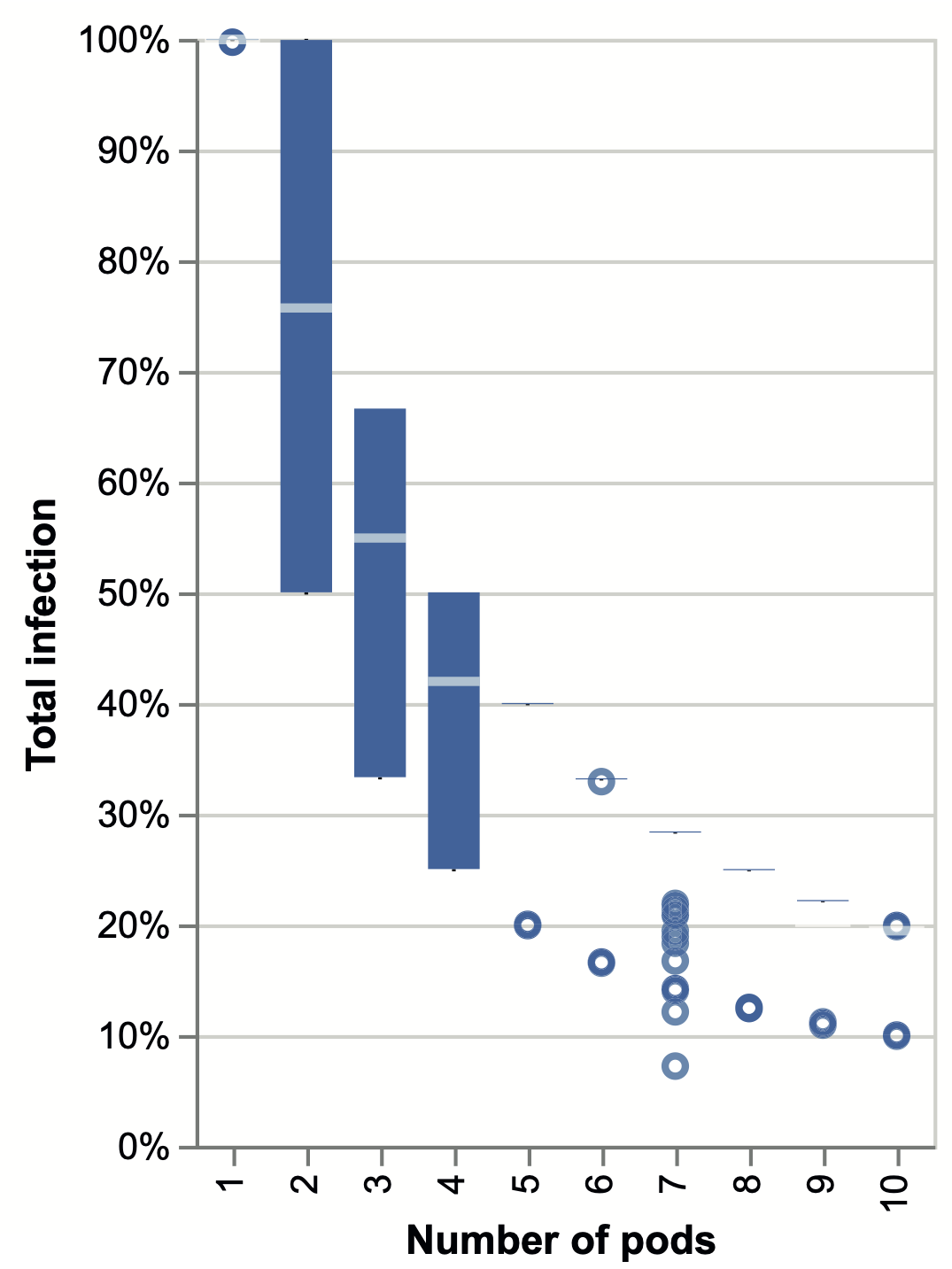} 
    \caption{Spatial pods, 2 patients zero} 
    \label{fig:sa} 
  \end{subfigure}
  \begin{subfigure}[b]{0.48\linewidth}
    \centering
    \includegraphics[scale=0.68]{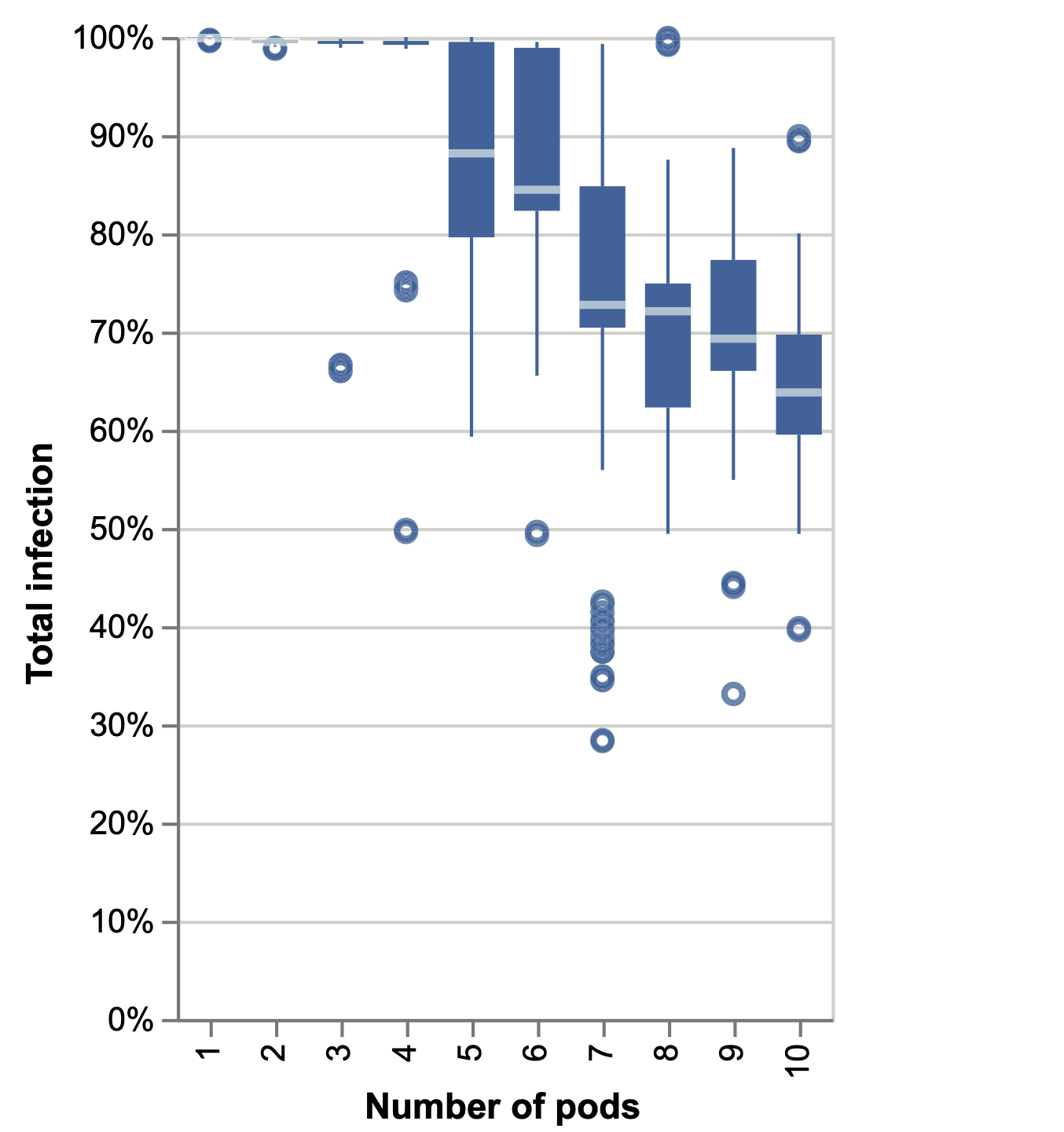}
    \caption{Spatial pods, 10 patients zero}
    \label{fig:sc} 
  \end{subfigure}
  \caption{The percentage of the population infected with an interaction-driven model over temporal random networks with spatial pods. The reduced \textit{spatial} density is obtained by separation to disjoint groups (the x-axis denotes the number of groups). Left panel (a) depicts infection with two randomly selected patients zero, emulating the early stages of the pandemic; Right panel (b) depicts infection with ten randomly selected patients zero, emulating a situation in which the disease is prevalent.   Each boxplot depicts the results of 200 corresponding simulator iterations over the 10000 temporal networks. Thus each graph is the result of 2000 iterations. The results show that spatial pods are more helpful at the early stages of the disease.}
  \label{fig:split} 
\end{figure}

To examine the effectiveness of \textit{spatial} pods, we conduct an experiment in which we simulate graphs consisting of completely disconnected sub-graphs, with the initial patients (patients zero) distributed between the sub-graphs at random.. Starting with our initial configuration of $1000$ nodes, we created sequences of subgraphs of a size defined by the required split. For example, we created two sequences of subgraphs of 500 nodes each for the period necessary to create the split into two disjoint groups.

Figure~\ref{fig:split} shows the results of using social distancing pods of various sizes by creating splits into two to ten disjoint groups. Figure~\ref{fig:sa} shows the effectiveness of social distancing pods in the early days of the disease when only a small number of people can infect. Specifically, we chose two initial patients randomly placed on the first day. Figure~\ref{fig:sc} depicts the results for the same experiments only with ten initial patients that were randomly placed.

\subsection*{Real-world networks with dilated temporal activities}
\label{sec:realw}
Considering the spatial pods' weaker effect at more advanced stages of the disease, we move on to examining the effect of dilation of the \textit{temporal} activity timeline on the spread of the disease. We dilate the timeline while maintaining the activity potential by spreading the activity over longer periods. As noted before, while it reduces the temporal density, it prolongs the period in which meetings occur, creating other opportunities for contagion. 

Here, we examine the effect of timeline dilation on a network of real-world encounters rather than a random network. The CNS temporal network, devised from proximity data, is amongst the detailed available contact data collected during the pre-pandemic era (see Supplementary Section 2.1).
 
\begin{figure}[!ht]
\centering
 \begin{subfigure}[t]{0.3\textwidth}
    \includegraphics[width=\textwidth]{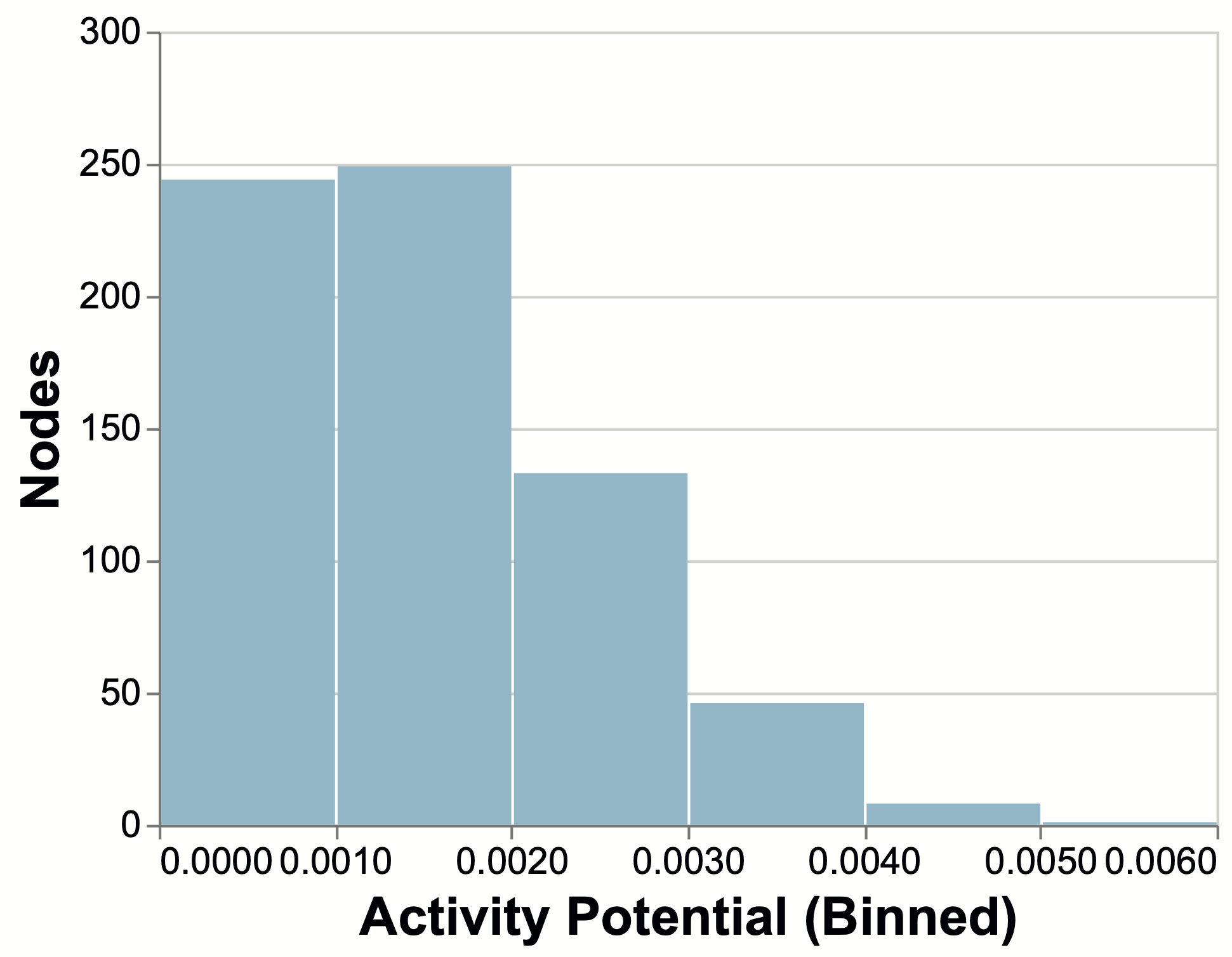} 
    \caption{28 daily time windows} 
    \label{fig:tl1} 
  \end{subfigure}
   \begin{subfigure}[t]{0.3\textwidth}
    \includegraphics[width=\textwidth]{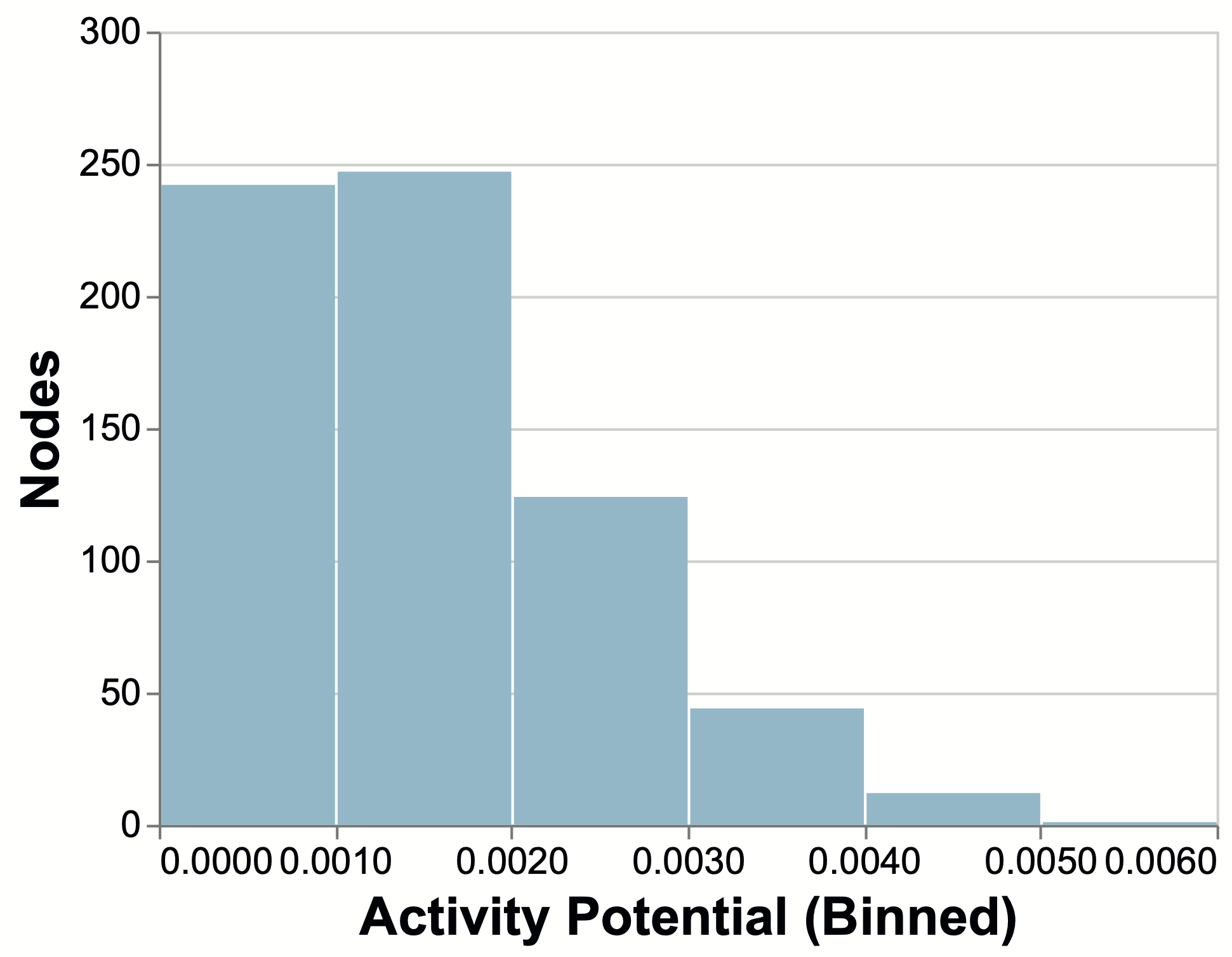} 
    \caption{28 half-day time windows} 
    \label{fig:tl2} 
  \end{subfigure}
   \begin{subfigure}[t]{0.3\textwidth}
    \includegraphics[width=\textwidth]{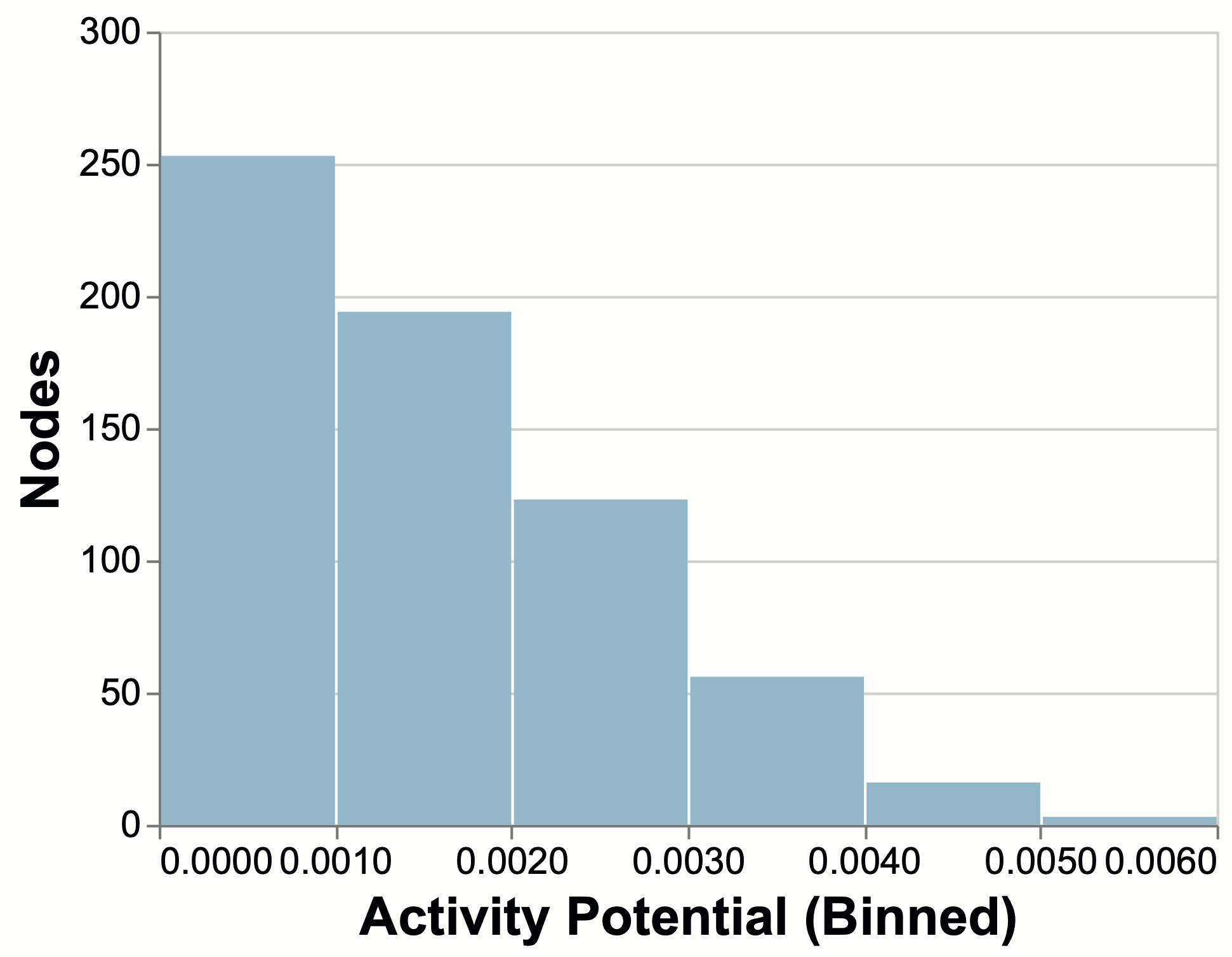} 
    \caption{28 one-third of a day time windows} 
    \label{fig:tl3} 
  \end{subfigure}

\caption{Histogram of activity potential in the CNS social networks for various time windows. X-axis is the activity potential per node, binned, whereas y-axis denotes the amount of nodes in the bin. The activity potential is preserved when observing different sized time-windows: (a) 28 $\times$ daily time windows (all data); (b) 28 $\times$ half-day time-windows, each time-window contains the interactions of half of the original time window; (c) 28 $\times$ one-third of a day time windows, each time-window contains the interactions of one-third of the original time-window.}
\label{fig:tl}
 \end{figure}

In this experiment, we create temporal pods by diluting the overall interactions in each time window. We split the interactions in each time window into  $1/k, k \in [2,3]$ of the original day while maintaining spatial and temporal ordering and the full activity of each node along the timeline. Thus, for example, a network dilated to half the daily interactions will be depicted as twice as long, time-window-wise.  To understand the difference per day, we show in Figure~\ref{fig:tl} the actual activity histogram of the network per 28-time windows of different lengths. 

\begin{figure}[!ht]
\centering
\vspace{-1cm}
 \begin{subfigure}[t]{.8\textwidth}
    \includegraphics[width=\textwidth]{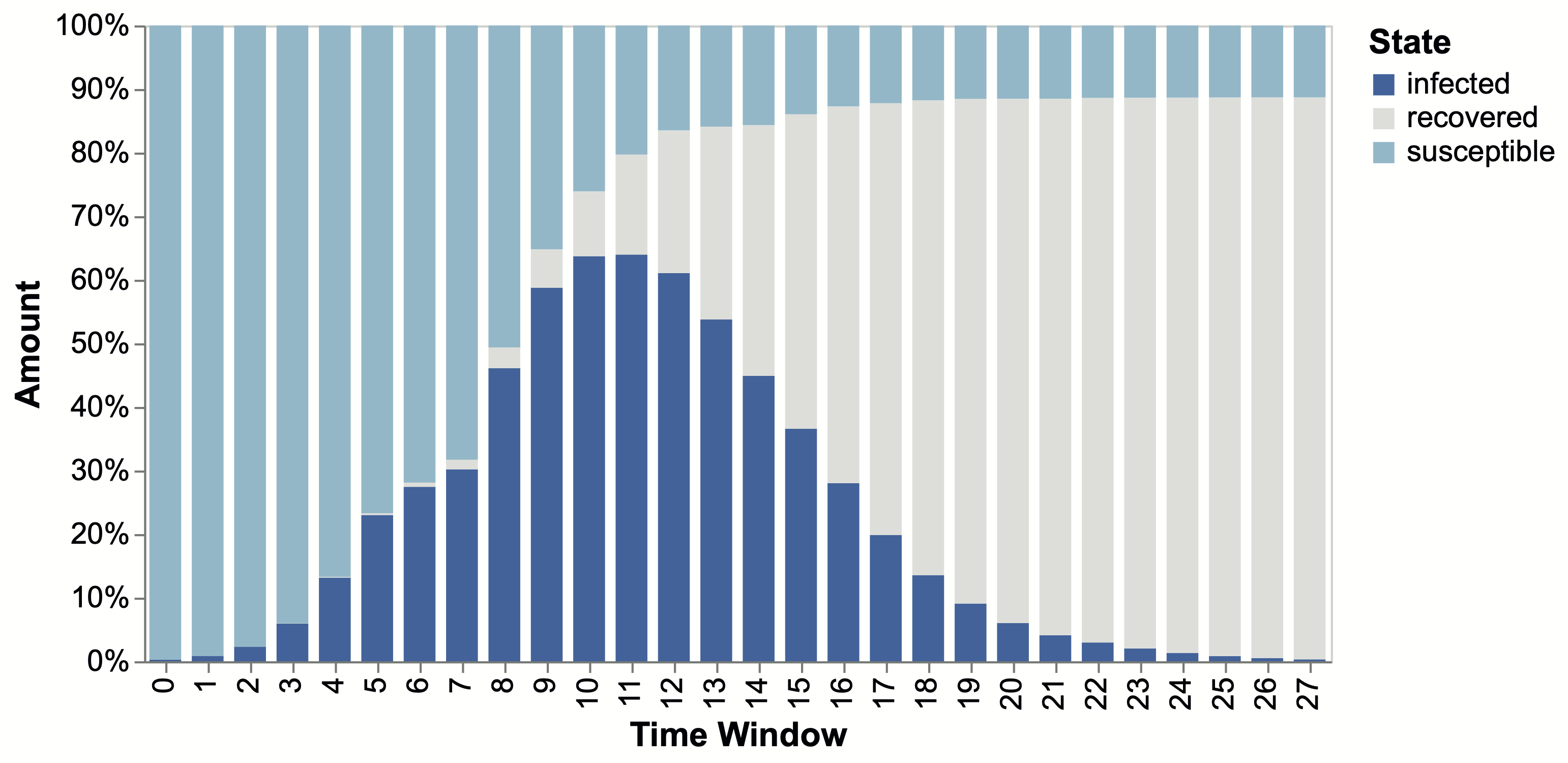} 
    \caption{Original network} 
    \label{fig:rwd1} 
    \vspace{1ex}
    \vspace{-.2cm}
  \end{subfigure}
  \begin{subfigure}[t]{0.8\textwidth}
    \includegraphics[width=\textwidth]{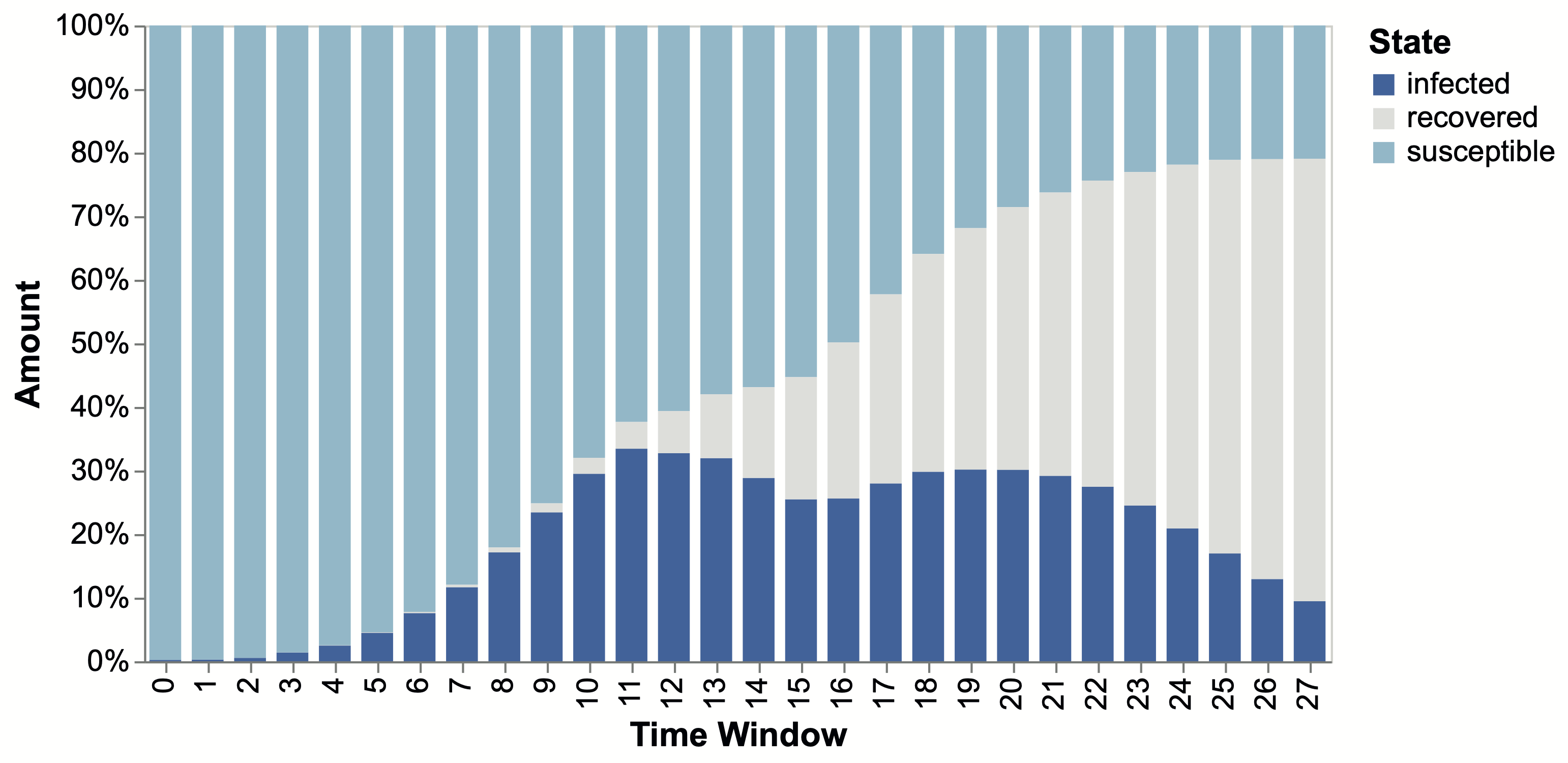} 
    \caption{Half temp. density} 
    \label{fig:rwd2} 
    \vspace{1ex}
    \vspace{-.2cm}
  \end{subfigure} 
  \begin{subfigure}[t]{0.8\textwidth}
    \includegraphics[width=\textwidth]{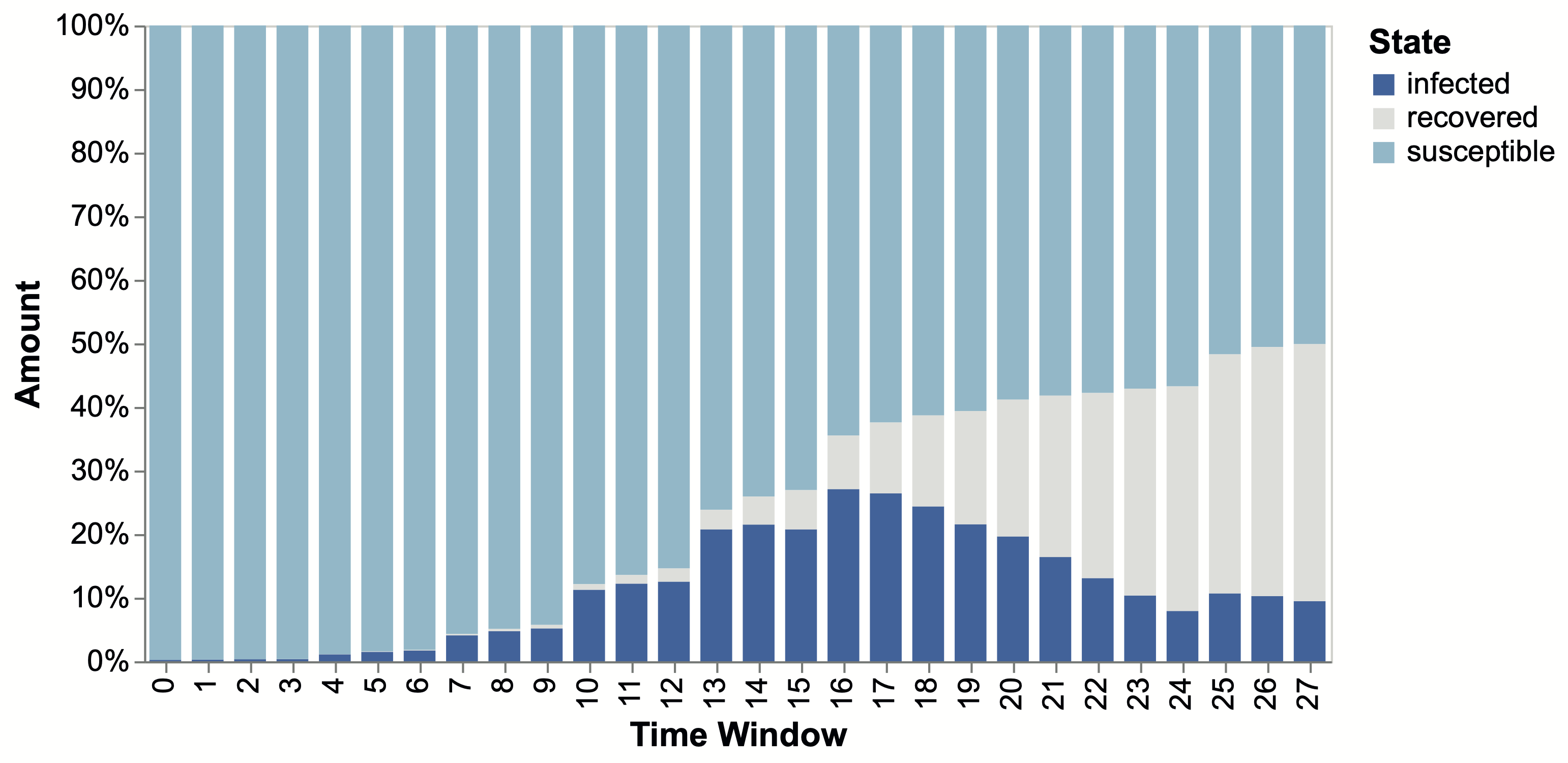} 
      \vspace{-.2cm}
    \caption{One third temp. density} 
    \label{fig:rwd3} 
  \end{subfigure}
  \vspace{-.1cm}
\caption{Disease spread in identical time frames while dilating temporal activity on the real-world CNS network. Only the first 29 days are presented.   (a) Original temporal network; (b) Network dilated to half the daily activity; (c) Network dilated to one-third daily activity.}
\label{fig:rwd}
\end{figure}

We examine our temporal interaction-driven contagion model over the resulted network. Each point results from 200 iterations with a randomly chosen initial patient zero. In the basic scenario, each day corresponds to 24 hours of activity. In the dilated networks, each time window is of length corresponding to the amount of dilation. For example, in networks with half a dilation, each time window consists of an interval of 12 hours of activities that are stretched over 24 hours. Similarly, with one-third dilation, each time window consists of an interval of 8 hours of activities that are stretched over 24 hours. In each of the encounters with infected individuals for which the minimal exposure time for infection is met or exceeded, the probability for infection also depends on $P_{\text{max}}$, as described in Equation~\ref{eq:vl2}. Here, we the required minimal exposure time to its minimum value, thus considering all interactions in the contagion process. To enable a visual comparison, we show the same number of time windows in the results depicted in Figure~\ref{fig:rwd}. When the interaction-driven contagion model is run over the original network, as seen in figure~\ref{fig:rwd1}, it takes less than 12 days to infect 80\% of the population, and by the end of the 29 days timeline, over 90\% of the population is infected. However, when the timeline is dilated, as in Figures~\ref{fig:rwd2} and~\ref{fig:rwd3}, the spread is slower, and the overall number of infected individuals is lower. 

\subsubsection*{The effect of timeline dilation on the contagion process of different variants}
\begin{figure}[!ht]
\centering
 \begin{subfigure}[t]{0.3\textwidth}
    \includegraphics[width=\textwidth]{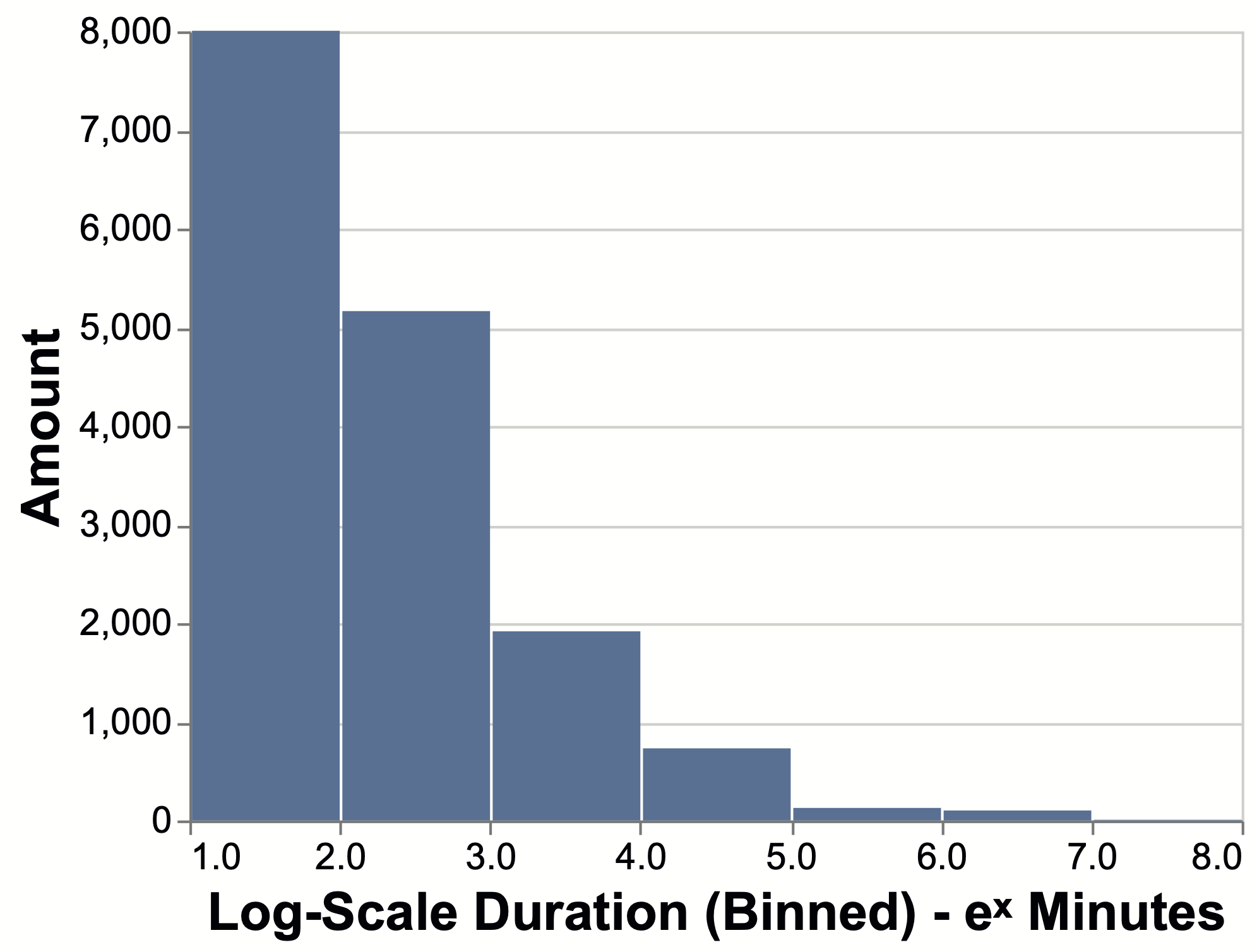} 
    \caption{24 hours intervals} 
    \label{fig:td1} 
  \end{subfigure}
   \begin{subfigure}[t]{0.3\textwidth}
    \includegraphics[width=\textwidth]{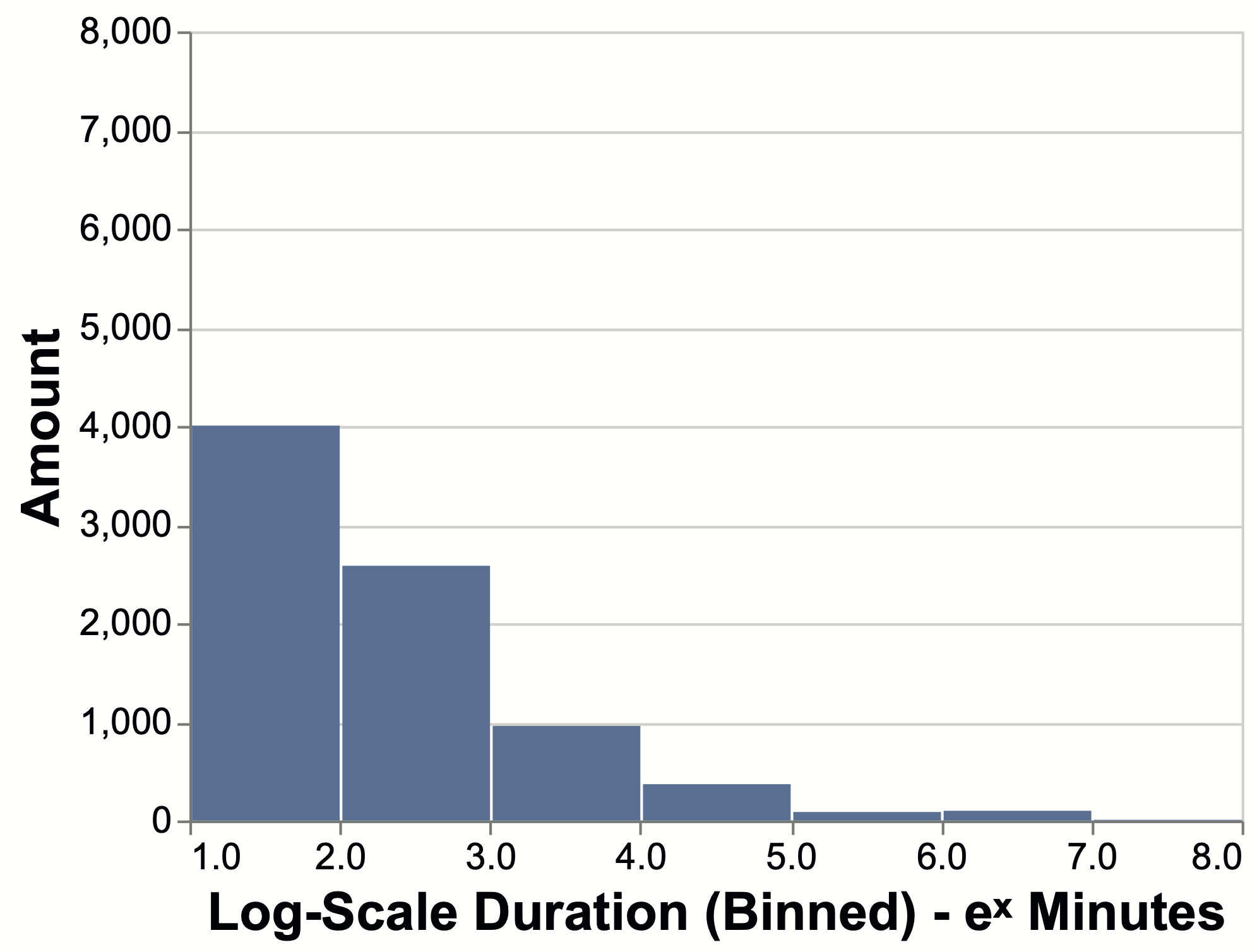} 
    \caption{Dilution in half} 
    \label{fig:td2} 
  \end{subfigure}
   \begin{subfigure}[t]{0.3\textwidth}
    \includegraphics[width=\textwidth]{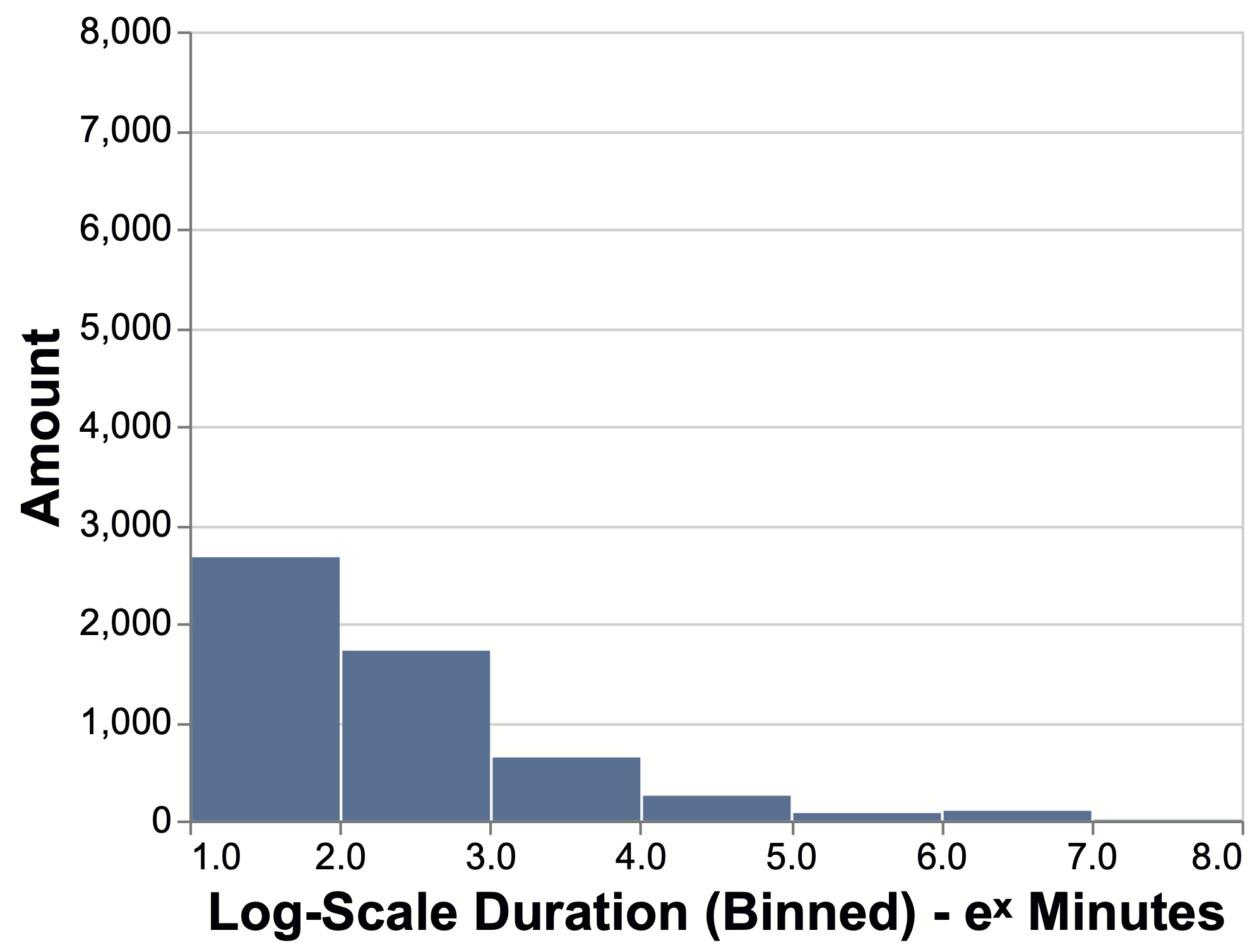} 
    \caption{Dilution by one-third} 
    \label{fig:td3} 
  \end{subfigure}
\caption{Order preserving timeline dilation: temporal meetings duration histogram of the CNS social networks. X-axis depicts a log-scale measure of meetings' duration,  y-axis denotes the amount: (a) Average daily amount of meetings of various duration in the CNS network measured over 24-hour intervals; (b) Average daily amount of meetings of various duration in the CNS network measured over 12-hour intervals. That is, each time window contains half the original interactions; (c) Average daily amount of meetings of various duration in the CNS network measured over 8-hour intervals. That is, each time window contains one-third the interactions of a full day.}
\label{fig:td}
\end{figure}

 Here, we also consider the meeting duration and various variants with different infection speeds. We correlate the infection speed of a variant with the minimal amount of time needed for infection, $D_{\text{min}}$. Figure~\ref{fig:td1} depicts the distribution of daily meeting lengths in the CNS dataset averaged over all days. The vast majority of the meetings are short, with a few meetings lasting longer than 200 minutes. Figures~\ref{fig:td2} and~\ref{fig:td3} show the corresponding amount of meetings of various duration in the network that is dilated to half and a third of the activity per time window.

We vary the minimal duration that correlates with infection, $D_{\text{min}}$ in the range $[5,120]$, by five minutes at each iteration. $D_{\text{min}}=5$ indicates a variant that is transmissible in all meetings, while $D_{\text{min}}=120$ corresponds to a variant that is transmissible only if exposure is very long. In each of the encounters with infected individuals for which the minimal exposure time for infection is met or exceeded, the probability for infection also depends on $P_{\text{max}}$, as described in Equation~\ref{eq:vl2}.

\begin{figure}[!ht]
\centering
\vspace{-1cm}
 \begin{subfigure}[t]{.7\textwidth}
    \includegraphics[width=\textwidth]{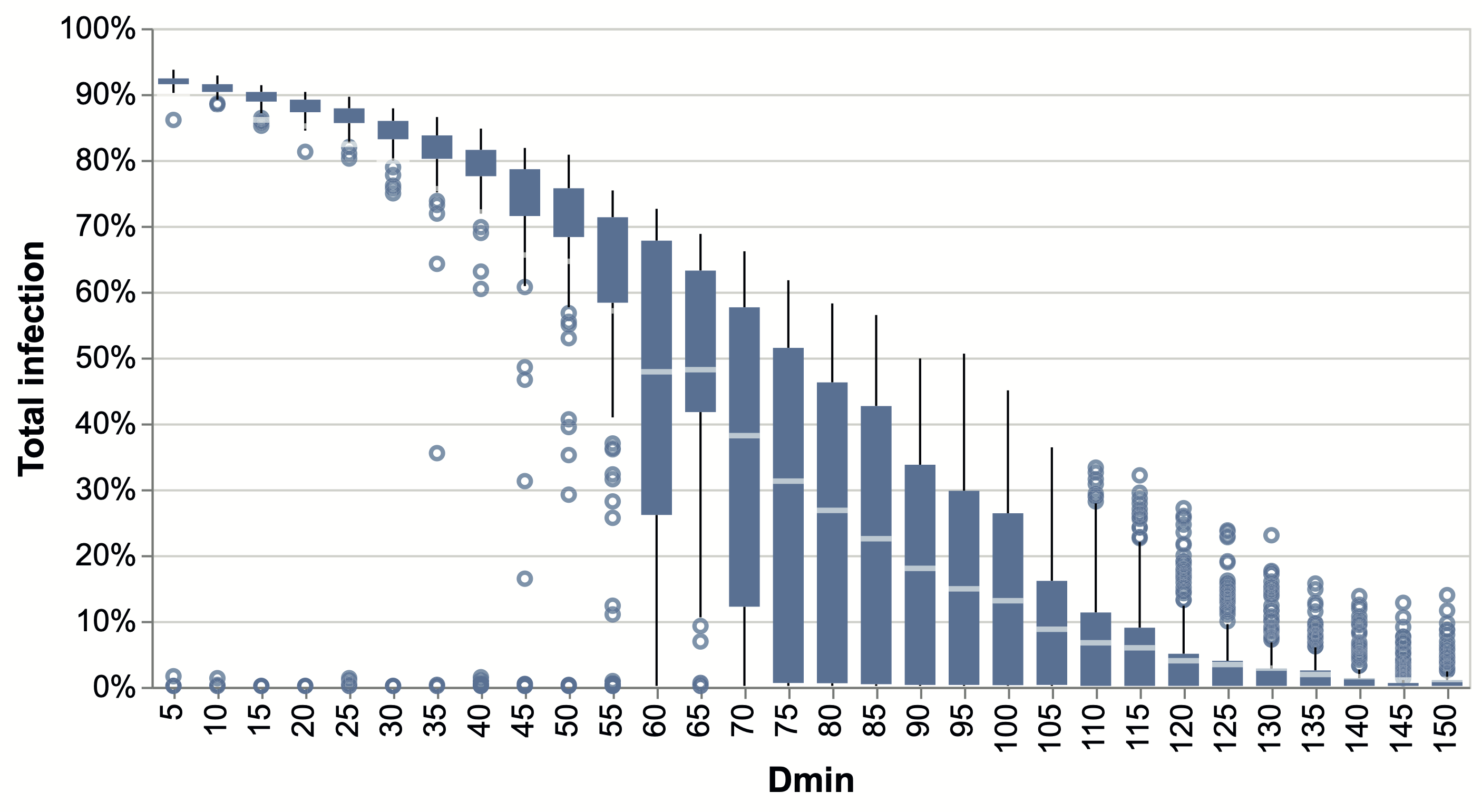} 
    \caption{Infection rate per $D_{\text{min}}$ - original network} 
    \label{fig:v1} 
    \vspace{1ex}
    \vspace{-.2cm}
  \end{subfigure}
  \begin{subfigure}[t]{0.7\textwidth}
    \includegraphics[width=\textwidth]{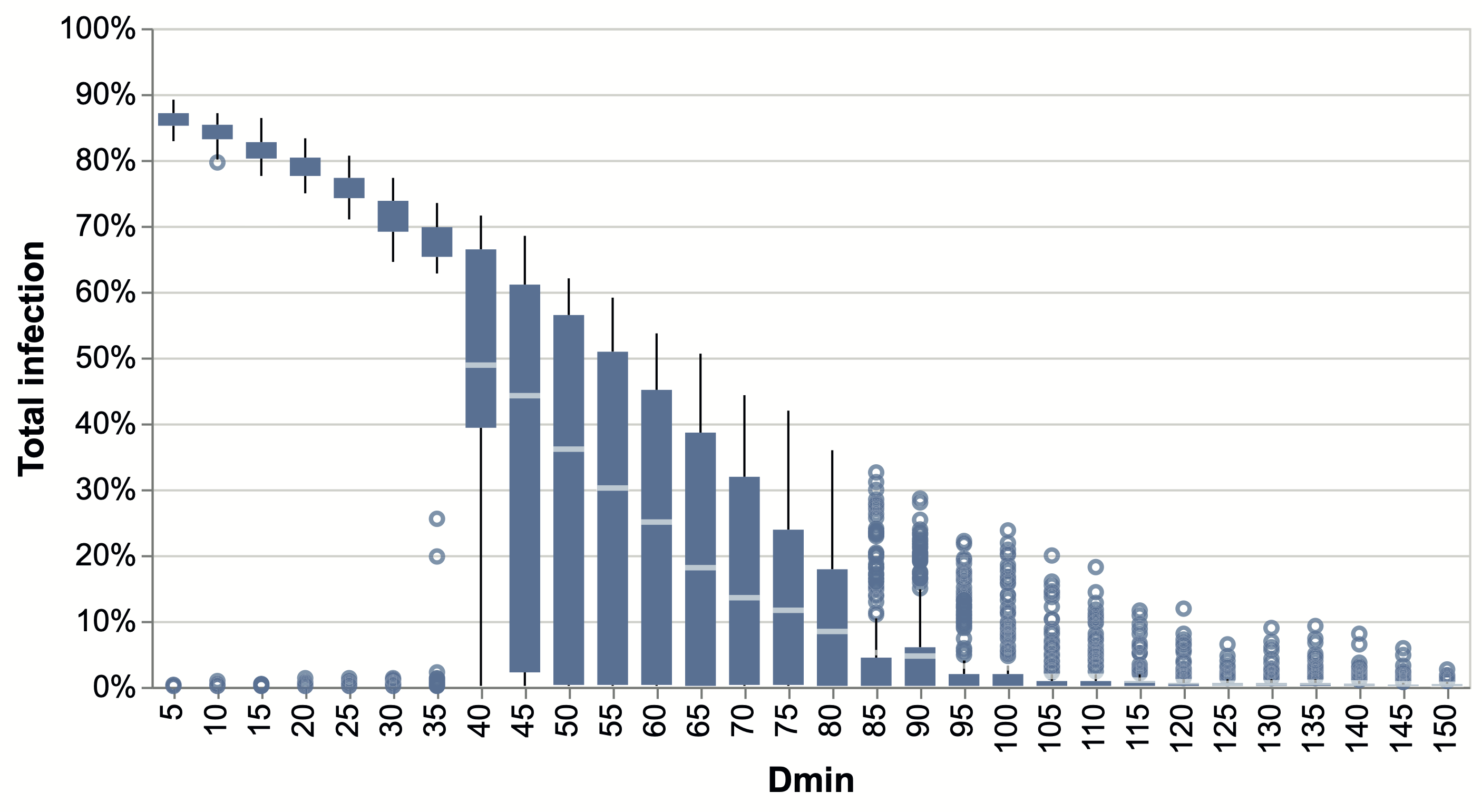} 
    \caption{Infection rate per $D_{\text{min}}$ - half day time window} 
    \label{fig:v2} 
    \vspace{1ex}
    \vspace{-.2cm}
  \end{subfigure} 
  \begin{subfigure}[t]{0.7\textwidth}
    \includegraphics[width=\textwidth]{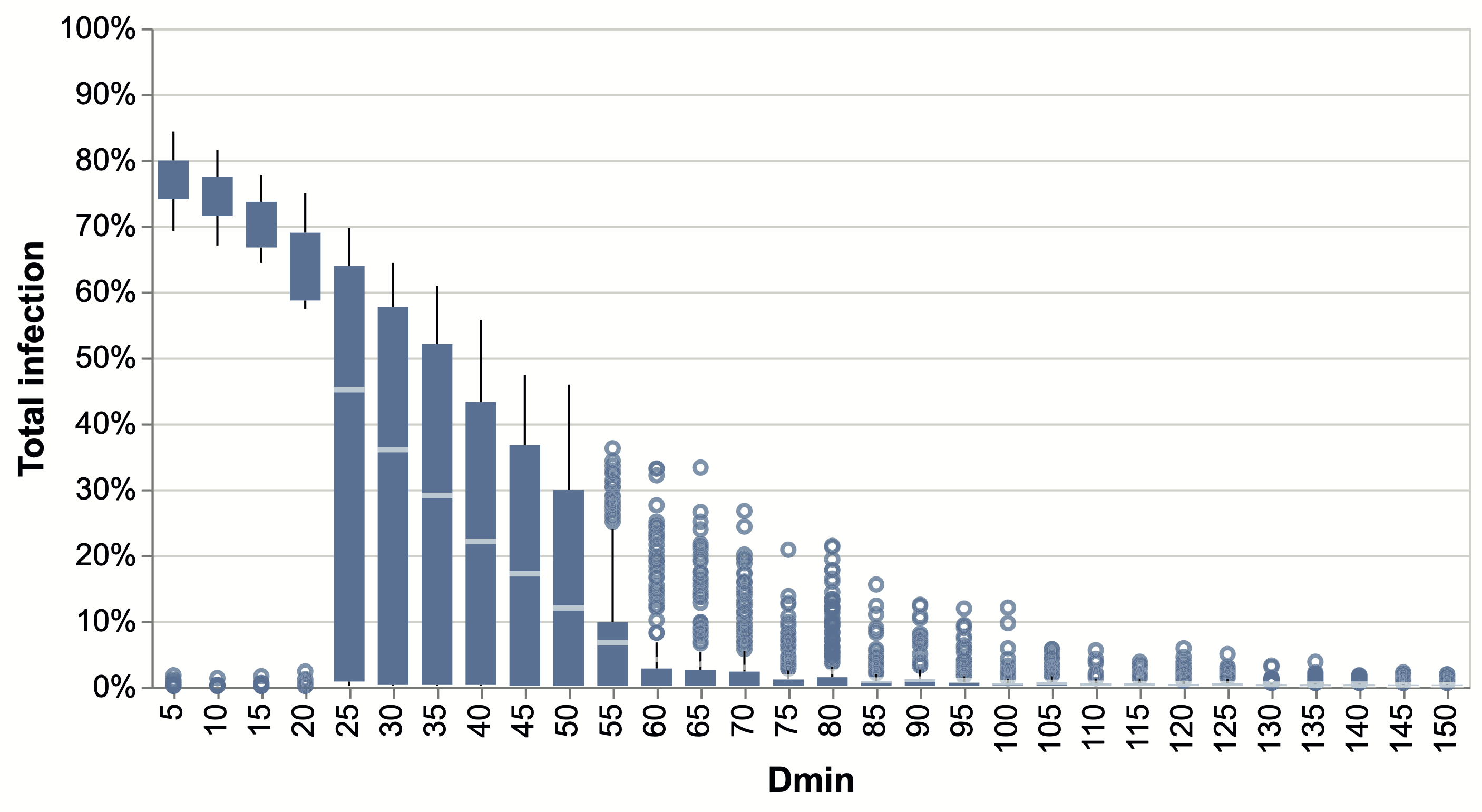} 
      \vspace{-.2cm}
    \caption{Infection rate per $D_{\text{min}}$ - one third of a day time window} 
    \label{fig:v3} 
  \end{subfigure}
  \vspace{-.1cm}
\caption{
The effect of timeline dilation on the infection rate for various variants that differ by the minimal exposure times needed for contagion (different $D_{\text{min}}$ values).  (a) Original  network; (b) Dilated  - network with half day time window; (c) Dilated  - network with one-third of day time window}
\label{fig:ve}
\end{figure}
Figure~\ref{fig:ve} depicts the spread of various pathogens requiring different minimal exposure values for contagion, corresponding to  various $D_{\text{min}}$ values over the CNS dataset. For each $D_{\text{min}}$ value, we performed 200 iterations with random placement of patient zero. The total infection rate is presented as a box-plot form of the distribution of the results measured over the 200 iterations for each of the $D_{\text{min}}$ values. We then experimented on the CNS dataset with dilated timelines - using time windows corresponding to half days and one-third of a day. Figure~\ref{fig:v1} shows the total infection rate for the different pathogens on the original network. Faster pathogens that need less exposure time for transmission infect most of the population. Slower pathogens that require higher exposure time to infect are less successful. However, as can be seen in Figures~\ref{fig:v2} and \ref{fig:v3}, \textit{dilation of the timeline lowers the ability of pathogens to infect}, as infected people have fewer opportunities to infect before they are removed. For example, a pathogen that spread to over 75\% of the population in the original network would infect less than 50\% of the community in a network with a dilated timeline and half the activity per time window. Moreover, In the original network, slow-spreading pathogens are almost as contagious as pathogens that are ten times as fast. In contrast, in the dilated timeline, slower pathogens do not spread as fast ones.

\section*{Discussion}

Temporal networks enable the research of disease progression while following possible routes the disease follows. This research departs from an aggregated view and takes a temporal path-respecting one and follows the spread of the disease over the timeline of interactions. 

We presented a realistic interaction-driven airborne disease model adapted to the COVID-19 known characteristics and showed that it could be used with randomly generated networks and real-life contact information. Our interaction-driven contagion model enables the consideration of meetings' duration. We model the duration as a proxy for the transmitted viral load~\cite{rea2007duration,smieszek2009mechanistic}. 
 Thus, our model can be used to model the spread of various variants~\cite{abbey2022analysis}.

As our model is tailored to work with temporal contact information, we work with two data types. Temporal random networks were generated according to the algorithm in~\cite{zhang2017random}, and the real-world contact network data from the Copenhagen Network Study~\cite{sapiezynski2019interaction}. Our package for generating temporal random networks, RandomDynamicGraph (RDG), is freely available at \url{https://github.com/ScanLab-ossi/DynamicRandomGraphs}.  

We used our interaction-driven model to assess the effectiveness of various social distancing policies over temporal random and real-world social networks. We model temporal pods by dilating the timeline while preserving the activity potential, as defined by Perra \etal~\cite{perra2012activity}. Social distancing (spatial) pods are modeled by dividing groups into disjoint subgroups. To model a spatiotemporal social distancing policy, we divide the network into two spatial pods that interact on alternate days. 
 
Our results demonstrated that spatial pods are effective with a low number of initial patients, which may correspond to a low transmissible epidemic or the early days of a more contagion one. Thus, applying social distancing pods is probably beneficial only at the early stages of a pandemic or for slowly transmitted diseases. In our simulations, the spatial pods are separated. In real life, however, such separation is often not feasible over time, which will further reduce the effectiveness of this solution. 
Spatial pods, however, can be used as a model for nonrandom mixing in populations, for example, when people with similar characteristics group together. Reports of nonrandom mixing in disease spreading exist for the measles outbreak in Chicago in the 1980s and the  Swine flu in Fort Dix in 1976~\cite{silver2012signal}. In these cases, the nonrandom mixing between the groups resulted in only one of the groups getting the disease. Nonrandom mixing patterns and their effect on spatial and temporal pods are part of our future work plan. 

Working with real-world contact networks, we observed that when the typical activity potential in the network is high, slower and, faster pathogens are almost as contagious. This is a surprising result, as slower pathogens follow different routes than faster ones since they cannot infect during short meetings. As most meetings are short, this is a surprising result. 
We further found that temporal pods, that is, dilated timelines, are quite effective in slowing down the spread of the disease and create a clear distinction between the infection rate of fast pathogens and slower ones. 

We find that the social dynamics are the key to the spread of the disease. These dynamics could be more dominant than the pathogen's specific characteristics in determining its progress in the community.   Our results indicate that social distancing policies~\cite{courtemanche2020strong} should reduce the activity potential rather than create spatial pods.

\bibliography{temporal}


\begin{thebibliography}{32}
\ifx \bisbn   \undefined \def \bisbn  #1{ISBN #1}\fi
\ifx \binits  \undefined \def \binits#1{#1}\fi
\ifx \bauthor  \undefined \def \bauthor#1{#1}\fi
\ifx \batitle  \undefined \def \batitle#1{#1}\fi
\ifx \bjtitle  \undefined \def \bjtitle#1{#1}\fi
\ifx \bvolume  \undefined \def \bvolume#1{\textbf{#1}}\fi
\ifx \byear  \undefined \def \byear#1{#1}\fi
\ifx \bissue  \undefined \def \bissue#1{#1}\fi
\ifx \bfpage  \undefined \def \bfpage#1{#1}\fi
\ifx \blpage  \undefined \def \blpage #1{#1}\fi
\ifx \burl  \undefined \def \burl#1{\textsf{#1}}\fi
\ifx \doiurl  \undefined \def \doiurl#1{\url{https://doi.org/#1}}\fi
\ifx \betal  \undefined \def \betal{\textit{et al.}}\fi
\ifx \binstitute  \undefined \def \binstitute#1{#1}\fi
\ifx \binstitutionaled  \undefined \def \binstitutionaled#1{#1}\fi
\ifx \bctitle  \undefined \def \bctitle#1{#1}\fi
\ifx \beditor  \undefined \def \beditor#1{#1}\fi
\ifx \bpublisher  \undefined \def \bpublisher#1{#1}\fi
\ifx \bbtitle  \undefined \def \bbtitle#1{#1}\fi
\ifx \bedition  \undefined \def \bedition#1{#1}\fi
\ifx \bseriesno  \undefined \def \bseriesno#1{#1}\fi
\ifx \blocation  \undefined \def \blocation#1{#1}\fi
\ifx \bsertitle  \undefined \def \bsertitle#1{#1}\fi
\ifx \bsnm \undefined \def \bsnm#1{#1}\fi
\ifx \bsuffix \undefined \def \bsuffix#1{#1}\fi
\ifx \bparticle \undefined \def \bparticle#1{#1}\fi
\ifx \barticle \undefined \def \barticle#1{#1}\fi
\bibcommenthead
\ifx \bconfdate \undefined \def \bconfdate #1{#1}\fi
\ifx \botherref \undefined \def \botherref #1{#1}\fi
\ifx \url \undefined \def \url#1{\textsf{#1}}\fi
\ifx \bchapter \undefined \def \bchapter#1{#1}\fi
\ifx \bbook \undefined \def \bbook#1{#1}\fi
\ifx \bcomment \undefined \def \bcomment#1{#1}\fi
\ifx \oauthor \undefined \def \oauthor#1{#1}\fi
\ifx \citeauthoryear \undefined \def \citeauthoryear#1{#1}\fi
\ifx \endbibitem  \undefined \def \endbibitem {}\fi
\ifx \bconflocation  \undefined \def \bconflocation#1{#1}\fi
\ifx \arxivurl  \undefined \def \arxivurl#1{\textsf{#1}}\fi
\csname PreBibitemsHook\endcsname

\bibitem{anderson2020will}
\begin{barticle}
\bauthor{\bsnm{Anderson}, \binits{R.M.}},
\bauthor{\bsnm{Heesterbeek}, \binits{H.}},
\bauthor{\bsnm{Klinkenberg}, \binits{D.}},
\bauthor{\bsnm{Hollingsworth}, \binits{T.D.}}:
\batitle{How will country-based mitigation measures influence the course of the
  covid-19 epidemic?}
\bjtitle{The lancet}
\bvolume{395}(\bissue{10228}),
\bfpage{931}--\blpage{934}
(\byear{2020})
\end{barticle}
\endbibitem

\bibitem{courtemanche2020strong}
\begin{barticle}
\bauthor{\bsnm{Courtemanche}, \binits{C.}},
\bauthor{\bsnm{Garuccio}, \binits{J.}},
\bauthor{\bsnm{Le}, \binits{A.}},
\bauthor{\bsnm{Pinkston}, \binits{J.}},
\bauthor{\bsnm{Yelowitz}, \binits{A.}}:
\batitle{Strong social distancing measures in the united states reduced the
  covid-19 growth rate: Study evaluates the impact of social distancing
  measures on the growth rate of confirmed covid-19 cases across the united
  states.}
\bjtitle{Health Affairs}
\bvolume{39}(\bissue{7}),
\bfpage{1237}--\blpage{1246}
(\byear{2020})
\end{barticle}
\endbibitem

\bibitem{flaxman2020estimating}
\begin{barticle}
\bauthor{\bsnm{Flaxman}, \binits{S.}},
\bauthor{\bsnm{Mishra}, \binits{S.}},
\bauthor{\bsnm{Gandy}, \binits{A.}},
\bauthor{\bsnm{Unwin}, \binits{H.J.T.}},
\bauthor{\bsnm{Mellan}, \binits{T.A.}},
\bauthor{\bsnm{Coupland}, \binits{H.}},
\bauthor{\bsnm{Whittaker}, \binits{C.}},
\bauthor{\bsnm{Zhu}, \binits{H.}},
\bauthor{\bsnm{Berah}, \binits{T.}},
\bauthor{\bsnm{Eaton}, \binits{J.W.}}, \betal:
\batitle{Estimating the effects of non-pharmaceutical interventions on covid-19
  in europe}.
\bjtitle{Nature}
\bvolume{584}(\bissue{7820}),
\bfpage{257}--\blpage{261}
(\byear{2020})
\end{barticle}
\endbibitem

\bibitem{ferguson2006strategies}
\begin{barticle}
\bauthor{\bsnm{Ferguson}, \binits{N.M.}},
\bauthor{\bsnm{Cummings}, \binits{D.A.}},
\bauthor{\bsnm{Fraser}, \binits{C.}},
\bauthor{\bsnm{Cajka}, \binits{J.C.}},
\bauthor{\bsnm{Cooley}, \binits{P.C.}},
\bauthor{\bsnm{Burke}, \binits{D.S.}}:
\batitle{Strategies for mitigating an influenza pandemic}.
\bjtitle{Nature}
\bvolume{442}(\bissue{7101}),
\bfpage{448}--\blpage{452}
(\byear{2006})
\end{barticle}
\endbibitem

\bibitem{donohue2020covid}
\begin{barticle}
\bauthor{\bsnm{Donohue}, \binits{J.M.}},
\bauthor{\bsnm{Miller}, \binits{E.}}:
\batitle{Covid-19 and school closures}.
\bjtitle{Jama}
\bvolume{324}(\bissue{9}),
\bfpage{845}--\blpage{847}
(\byear{2020})
\end{barticle}
\endbibitem

\bibitem{heymann2004influence}
\begin{barticle}
\bauthor{\bsnm{Heymann}, \binits{A.}},
\bauthor{\bsnm{Chodick}, \binits{G.}},
\bauthor{\bsnm{Reichman}, \binits{B.}},
\bauthor{\bsnm{Kokia}, \binits{E.}},
\bauthor{\bsnm{Laufer}, \binits{J.}}:
\batitle{Influence of school closure on the incidence of viral respiratory
  diseases among children and on health care utilization}.
\bjtitle{The Pediatric infectious disease journal}
\bvolume{23}(\bissue{7}),
\bfpage{675}--\blpage{677}
(\byear{2004})
\end{barticle}
\endbibitem

\bibitem{perra2012activity}
\begin{barticle}
\bauthor{\bsnm{Perra}, \binits{N.}},
\bauthor{\bsnm{Gon{\c{c}}alves}, \binits{B.}},
\bauthor{\bsnm{Pastor-Satorras}, \binits{R.}},
\bauthor{\bsnm{Vespignani}, \binits{A.}}:
\batitle{Activity driven modeling of time varying networks}.
\bjtitle{Scientific reports}
\bvolume{2}(\bissue{1}),
\bfpage{1}--\blpage{7}
(\byear{2012})
\end{barticle}
\endbibitem

\bibitem{holme2012temporal}
\begin{barticle}
\bauthor{\bsnm{Holme}, \binits{P.}},
\bauthor{\bsnm{Saram{\"a}ki}, \binits{J.}}:
\batitle{Temporal networks}.
\bjtitle{Physics reports}
\bvolume{519}(\bissue{3}),
\bfpage{97}--\blpage{125}
(\byear{2012})
\end{barticle}
\endbibitem

\bibitem{delvenne2015diffusion}
\begin{barticle}
\bauthor{\bsnm{Delvenne}, \binits{J.-C.}},
\bauthor{\bsnm{Lambiotte}, \binits{R.}},
\bauthor{\bsnm{Rocha}, \binits{L.E.}}:
\batitle{Diffusion on networked systems is a question of time or structure}.
\bjtitle{Nature communications}
\bvolume{6}(\bissue{1}),
\bfpage{1}--\blpage{10}
(\byear{2015})
\end{barticle}
\endbibitem

\bibitem{masuda2017introduction}
\begin{bchapter}
\bauthor{\bsnm{Masuda}, \binits{N.}},
\bauthor{\bsnm{Holme}, \binits{P.}}:
\bctitle{Introduction to temporal network epidemiology}.
In: \bbtitle{Temporal Network Epidemiology},
pp. \bfpage{1}--\blpage{16}.
\bpublisher{Springer},
\blocation{Switzerland}
(\byear{2017})
\end{bchapter}
\endbibitem

\bibitem{zhang2017random}
\begin{barticle}
\bauthor{\bsnm{Zhang}, \binits{X.}},
\bauthor{\bsnm{Moore}, \binits{C.}},
\bauthor{\bsnm{Newman}, \binits{M.E.}}:
\batitle{Random graph models for dynamic networks}.
\bjtitle{The European Physical Journal B}
\bvolume{90}(\bissue{10}),
\bfpage{1}--\blpage{14}
(\byear{2017})
\end{barticle}
\endbibitem

\bibitem{rea2007duration}
\begin{barticle}
\bauthor{\bsnm{Rea}, \binits{E.}},
\bauthor{\bsnm{Lafleche}, \binits{J.}},
\bauthor{\bsnm{Stalker}, \binits{S.}},
\bauthor{\bsnm{Guarda}, \binits{B.}},
\bauthor{\bsnm{Shapiro}, \binits{H.}},
\bauthor{\bsnm{Johnson}, \binits{I.}},
\bauthor{\bsnm{Bondy}, \binits{S.}},
\bauthor{\bsnm{Upshur}, \binits{R.}},
\bauthor{\bsnm{Russell}, \binits{M.}},
\bauthor{\bsnm{Eliasziw}, \binits{M.}}:
\batitle{Duration and distance of exposure are important predictors of
  transmission among community contacts of ontario sars cases}.
\bjtitle{Epidemiology \& Infection}
\bvolume{135}(\bissue{6}),
\bfpage{914}--\blpage{921}
(\byear{2007})
\end{barticle}
\endbibitem

\bibitem{smieszek2009mechanistic}
\begin{barticle}
\bauthor{\bsnm{Smieszek}, \binits{T.}}:
\batitle{A mechanistic model of infection: why duration and intensity of
  contacts should be included in models of disease spread}.
\bjtitle{Theoretical Biology and Medical Modelling}
\bvolume{6}(\bissue{1}),
\bfpage{1}--\blpage{10}
(\byear{2009})
\end{barticle}
\endbibitem

\bibitem{stehle2011simulation}
\begin{barticle}
\bauthor{\bsnm{Stehl{\'e}}, \binits{J.}},
\bauthor{\bsnm{Voirin}, \binits{N.}},
\bauthor{\bsnm{Barrat}, \binits{A.}},
\bauthor{\bsnm{Cattuto}, \binits{C.}},
\bauthor{\bsnm{Colizza}, \binits{V.}},
\bauthor{\bsnm{Isella}, \binits{L.}},
\bauthor{\bsnm{R{\'e}gis}, \binits{C.}},
\bauthor{\bsnm{Pinton}, \binits{J.-F.}},
\bauthor{\bsnm{Khanafer}, \binits{N.}},
\bauthor{\bparticle{Van~den} \bsnm{Broeck}, \binits{W.}}, \betal:
\batitle{Simulation of an seir infectious disease model on the dynamic contact
  network of conference attendees}.
\bjtitle{BMC medicine}
\bvolume{9}(\bissue{1}),
\bfpage{87}
(\byear{2011})
\end{barticle}
\endbibitem

\bibitem{nagel2021realistic}
\begin{barticle}
\bauthor{\bsnm{Nagel}, \binits{K.}},
\bauthor{\bsnm{Rakow}, \binits{C.}},
\bauthor{\bsnm{M{\"u}ller}, \binits{S.A.}}:
\batitle{Realistic agent-based simulation of infection dynamics and
  percolation}.
\bjtitle{Physica A: Statistical Mechanics and its Applications}
\bvolume{584},
\bfpage{126322}
(\byear{2021})
\end{barticle}
\endbibitem

\bibitem{echeverria2021estimating}
\begin{barticle}
\bauthor{\bsnm{Echeverr{\'\i}a-Huarte}, \binits{I.}},
\bauthor{\bsnm{Garcimart{\'\i}n}, \binits{A.}},
\bauthor{\bsnm{Hidalgo}, \binits{R.}},
\bauthor{\bsnm{Mart{\'\i}n-G{\'o}mez}, \binits{C.}},
\bauthor{\bsnm{Zuriguel}, \binits{I.}}:
\batitle{Estimating density limits for walking pedestrians keeping a safe
  interpersonal distancing}.
\bjtitle{Scientific reports}
\bvolume{11}(\bissue{1}),
\bfpage{1}--\blpage{8}
(\byear{2021})
\end{barticle}
\endbibitem

\bibitem{stopczynski2014measuring}
\begin{botherref}
\oauthor{\bsnm{Stopczynski}, \binits{A.}},
\oauthor{\bsnm{Sekara}, \binits{V.}},
\oauthor{\bsnm{Sapiezynski}, \binits{P.}},
\oauthor{\bsnm{Cuttone}, \binits{A.}},
\oauthor{\bsnm{Madsen}, \binits{M.M.}},
\oauthor{\bsnm{Larsen}, \binits{J.E.}},
\oauthor{\bsnm{Lehmann}, \binits{S.}}:
Measuring large-scale social networks with high resolution.
PloS one
\textbf{9}(4)
(2014)
\end{botherref}
\endbibitem

\bibitem{Stopczynski:2015aa}
\begin{barticle}
\bauthor{\bsnm{Stopczynski}, \binits{A.}},
\bauthor{\bsnm{Sapiezynski}, \binits{P.}},
\bauthor{\bsnm{Pentland}, \binits{A.S.}},
\bauthor{\bsnm{Lehmann}, \binits{S.}}:
\batitle{Temporal fidelity in dynamic social networks}.
\bjtitle{The European Physical Journal B}
\bvolume{88}(\bissue{10}),
\bfpage{249}
(\byear{2015}).
\doiurl{10.1140/epjb/e2015-60549-7}
\end{barticle}
\endbibitem

\bibitem{sapiezynski2019interaction}
\begin{barticle}
\bauthor{\bsnm{Sapiezynski}, \binits{P.}},
\bauthor{\bsnm{Stopczynski}, \binits{A.}},
\bauthor{\bsnm{Lassen}, \binits{D.D.}},
\bauthor{\bsnm{Lehmann}, \binits{S.}}:
\batitle{Interaction data from the copenhagen networks study}.
\bjtitle{Scientific Data}
\bvolume{6}(\bissue{1}),
\bfpage{1}--\blpage{10}
(\byear{2019})
\end{barticle}
\endbibitem

\bibitem{barabasi2005origin}
\begin{barticle}
\bauthor{\bsnm{Barabasi}, \binits{A.-L.}}:
\batitle{The origin of bursts and heavy tails in human dynamics}.
\bjtitle{Nature}
\bvolume{435}(\bissue{7039}),
\bfpage{207}
(\byear{2005})
\end{barticle}
\endbibitem

\bibitem{mokryn2016role}
\begin{barticle}
\bauthor{\bsnm{Mokryn}, \binits{O.}},
\bauthor{\bsnm{Wagner}, \binits{A.}},
\bauthor{\bsnm{Blattner}, \binits{M.}},
\bauthor{\bsnm{Ruppin}, \binits{E.}},
\bauthor{\bsnm{Shavitt}, \binits{Y.}}:
\batitle{The role of temporal trends in growing networks}.
\bjtitle{PloS one}
\bvolume{11}(\bissue{8}),
\bfpage{0156505}
(\byear{2016})
\end{barticle}
\endbibitem

\bibitem{fumanelli2012inferring}
\begin{botherref}
\oauthor{\bsnm{Fumanelli}, \binits{L.}},
\oauthor{\bsnm{Ajelli}, \binits{M.}},
\oauthor{\bsnm{Manfredi}, \binits{P.}},
\oauthor{\bsnm{Vespignani}, \binits{A.}},
\oauthor{\bsnm{Merler}, \binits{S.}}:
Inferring the structure of social contacts from demographic data in the
  analysis of infectious diseases spread.
PLoS computational biology
\textbf{8}(9)
(2012)
\end{botherref}
\endbibitem

\bibitem{pastor2015epidemic}
\begin{barticle}
\bauthor{\bsnm{Pastor-Satorras}, \binits{R.}},
\bauthor{\bsnm{Castellano}, \binits{C.}},
\bauthor{\bsnm{Van~Mieghem}, \binits{P.}},
\bauthor{\bsnm{Vespignani}, \binits{A.}}:
\batitle{Epidemic processes in complex networks}.
\bjtitle{Reviews of modern physics}
\bvolume{87}(\bissue{3}),
\bfpage{925}
(\byear{2015})
\end{barticle}
\endbibitem

\bibitem{li2007role}
\begin{barticle}
\bauthor{\bsnm{Li}, \binits{Y.}},
\bauthor{\bsnm{Leung}, \binits{G.M.}},
\bauthor{\bsnm{Tang}, \binits{J.}},
\bauthor{\bsnm{Yang}, \binits{X.}},
\bauthor{\bsnm{Chao}, \binits{C.}},
\bauthor{\bsnm{Lin}, \binits{J.Z.}},
\bauthor{\bsnm{Lu}, \binits{J.}},
\bauthor{\bsnm{Nielsen}, \binits{P.V.}},
\bauthor{\bsnm{Niu}, \binits{J.}},
\bauthor{\bsnm{Qian}, \binits{H.}}, \betal:
\batitle{Role of ventilation in airborne transmission of infectious agents in
  the built environment-a multidisciplinary systematic review.}
\bjtitle{Indoor air}
\bvolume{17}(\bissue{1}),
\bfpage{2}--\blpage{18}
(\byear{2007})
\end{barticle}
\endbibitem

\bibitem{noakes2006modelling}
\begin{barticle}
\bauthor{\bsnm{Noakes}, \binits{C.}},
\bauthor{\bsnm{Beggs}, \binits{C.}},
\bauthor{\bsnm{Sleigh}, \binits{P.}},
\bauthor{\bsnm{Kerr}, \binits{K.}}:
\batitle{Modelling the transmission of airborne infections in enclosed spaces}.
\bjtitle{Epidemiology \& Infection}
\bvolume{134}(\bissue{5}),
\bfpage{1082}--\blpage{1091}
(\byear{2006})
\end{barticle}
\endbibitem

\bibitem{sze2010review}
\begin{barticle}
\bauthor{\bsnm{Sze~To}, \binits{G.N.}},
\bauthor{\bsnm{Chao}, \binits{C.Y.H.}}:
\batitle{Review and comparison between the wells--riley and dose-response
  approaches to risk assessment of infectious respiratory diseases}.
\bjtitle{Indoor air}
\bvolume{20}(\bissue{1}),
\bfpage{2}--\blpage{16}
(\byear{2010})
\end{barticle}
\endbibitem

\bibitem{ferretti2020quantifying}
\begin{botherref}
\oauthor{\bsnm{Ferretti}, \binits{L.}},
\oauthor{\bsnm{Wymant}, \binits{C.}},
\oauthor{\bsnm{Kendall}, \binits{M.}},
\oauthor{\bsnm{Zhao}, \binits{L.}},
\oauthor{\bsnm{Nurtay}, \binits{A.}},
\oauthor{\bsnm{Abeler-D{\"o}rner}, \binits{L.}},
\oauthor{\bsnm{Parker}, \binits{M.}},
\oauthor{\bsnm{Bonsall}, \binits{D.}},
\oauthor{\bsnm{Fraser}, \binits{C.}}:
Quantifying sars-cov-2 transmission suggests epidemic control with digital
  contact tracing.
Science
\textbf{368}(6491)
(2020)
\end{botherref}
\endbibitem

\bibitem{luo2021infection}
\begin{botherref}
\oauthor{\bsnm{Luo}, \binits{C.H.}},
\oauthor{\bsnm{Morris}, \binits{C.P.}},
\oauthor{\bsnm{Sachithanandham}, \binits{J.}},
\oauthor{\bsnm{Amadi}, \binits{A.}},
\oauthor{\bsnm{Gaston}, \binits{D.}},
\oauthor{\bsnm{Li}, \binits{M.}},
\oauthor{\bsnm{Swanson}, \binits{N.J.}},
\oauthor{\bsnm{Schwartz}, \binits{M.}},
\oauthor{\bsnm{Klein}, \binits{E.Y.}},
\oauthor{\bsnm{Pekosz}, \binits{A.}}, et al.:
Infection with the sars-cov-2 delta variant is associated with higher
  infectious virus loads compared to the alpha variant in both unvaccinated and
  vaccinated individuals.
medRxiv
(2021)
\end{botherref}
\endbibitem

\bibitem{teyssou2021delta}
\begin{barticle}
\bauthor{\bsnm{Teyssou}, \binits{E.}},
\bauthor{\bsnm{Delagr{\`e}verie}, \binits{H.}},
\bauthor{\bsnm{Visseaux}, \binits{B.}},
\bauthor{\bsnm{Lambert-Niclot}, \binits{S.}},
\bauthor{\bsnm{Brichler}, \binits{S.}},
\bauthor{\bsnm{Ferre}, \binits{V.}},
\bauthor{\bsnm{Marot}, \binits{S.}},
\bauthor{\bsnm{Jary}, \binits{A.}},
\bauthor{\bsnm{Todesco}, \binits{E.}},
\bauthor{\bsnm{Schnuriger}, \binits{A.}}, \betal:
\batitle{The delta sars-cov-2 variant has a higher viral load than the beta and
  the historical variants in nasopharyngeal samples from newly diagnosed
  covid-19 patients}.
\bjtitle{Journal of Infection}
\bvolume{83}(\bissue{4}),
\bfpage{1}--\blpage{3}
(\byear{2021})
\end{barticle}
\endbibitem

\bibitem{miller2020size}
\begin{barticle}
\bauthor{\bsnm{Miller}, \binits{H.}},
\bauthor{\bsnm{Mokryn}, \binits{O.}}:
\batitle{Size agnostic change point detection framework for evolving networks}.
\bjtitle{Plos one}
\bvolume{15}(\bissue{4}),
\bfpage{0231035}
(\byear{2020})
\end{barticle}
\endbibitem

\bibitem{abbey2022analysis}
\begin{barticle}
\bauthor{\bsnm{Abbey}, \binits{A.}},
\bauthor{\bsnm{Shahar}, \binits{Y.}},
\bauthor{\bsnm{Mokryn}, \binits{O.}}:
\batitle{Analysis of the competition among viral strains using a temporal
  interaction-driven contagion model}.
\bjtitle{Scientific Reports}
\bvolume{12}(\bissue{1}),
\bfpage{1}--\blpage{10}
(\byear{2022})
\end{barticle}
\endbibitem

\bibitem{silver2012signal}
\begin{bbook}
\bauthor{\bsnm{Silver}, \binits{N.}}:
\bbtitle{The Signal and the Noise: the Art and Science of Prediction}.
\bpublisher{Penguin UK},
\blocation{UK}
(\byear{2012})
\end{bbook}
\endbibitem

\end{thebibliography}
\omitit{

}

\section*{Author contributions}

O.M., Y.S., A.A., Y.M. designed the experiments; A.A. and Y.M. wrote the code and performed all the experiments; All authors analyzed the results; O.M. wrote the paper; All authors reviewed the manuscript.

Correspondence to Osnat Mokryn.

\section*{Additional information}
\subsection*{Competing interests
}

The authors declare no competing interests.

\end{document}